\documentclass[preprint,3p,10pt]{elsarticle}
\usepackage{lineno,hyperref}
\usepackage{amsmath,bm,geometry}
\usepackage{amsfonts}
\usepackage{color}
\usepackage{nomencl,etoolbox,ragged2e,siunitx}
\usepackage{mathtools}
\usepackage{multirow}
\usepackage{caption}
\usepackage{parskip}
\usepackage{calligra}
\usepackage{makeidx}
\usepackage{tikz}
\usepackage{mathrsfs}
\usepackage{bbm}
\usepackage{algorithm}
\usepackage{algorithmic}
\usepackage{amsthm}
\usepackage{graphicx}
\usepackage{subfigure}
\usepackage{hyperref}
\usepackage{cleveref}
\usepackage{setspace}
\hypersetup{pdfauthor=author}
\usetikzlibrary{matrix}
\makeindex
\DeclareMathAlphabet{\mathcalligra}{T1}{calligra}{m}{n}
\newtheorem{theorem}{Theorem}
\let\oldequation\equation
\let\oldendequation\endequation
\renewenvironment{equation}
{\linenomathNonumbers\oldequation}
{\oldendequation\endlinenomath}

\DeclareMathOperator{\arccot}{arccot}

\begin{document}
\title{An efficient numerical method for high energy $\alpha$ particle transport 
based on a hybrid collision model and machine learning}
\author[ad1]{Chang Liu}
\ead{liuchang@iapcm.ac.cn}
\author[ad1]{Bao Du}
\ead{du\_bao@iapcm.ac.cn}
\author[ad1,ad2]{Peng Song}
\ead{song\_peng@iapcm.ac.cn}
\address[ad1]{Institute of Applied Physics and Computational Mathematics, Beijing, China}
\address[ad2]{HEDPS, Center for Applied Physics and Technology, College of Engineering, Peking University, Beijing, China}
\cortext[cor1]{Corresponding author}

\begin{abstract}
The plasma heating by the $\alpha$ particle transport is 
a main self-heating source in inertial confinement fusion (ICF) 
that determines the capsule implosion performance. 
Due to the high energy of $\alpha$ particle and 
the high temperature in the ICF capsule hot spot, 
significant non-equilibrium effect exists and the continuum mechanics breaks down.
For the numerical simulation of implosion and charged particle transport, 
the Boltzmann equation needs to be solved to capture the kinetic effects.
However, the 7-dimensional Boltzmann equation, the highly frequent Coulomb collision, 
and the multi-folded integral stopping power formulation 
greatly limits the computational efficiency and challenges the computational power.
To overcome the high computational cost of high-frequent coulomb collisions, 
we propose a hybrid collision model according to which 
the collisions are categorized into low frequent large-angle collisions and 
high frequent small-angle grazing collisions.
The large-angle collision process is precisely solved based on the Coulomb cross-section.
For the highly frequent small-angle grazing, 
a statistic model is constructed 
with second-order accuracy in time.
The hybrid collision model reduces the computational cost of scattering calculation by two magnitudes.
For the multi-folded integral stopping power formulation, 
a neural network is used to improve computational efficiency.
Based on the proposed algorithm, 
we develop one-dimensional to three-dimensional module code 
to directly solve the the $\alpha$ particle transport Boltzmann equation. 
The $\alpha$ transport module code is integrated into the multi-physics LARED-S program.
The MC version ICF software is verified by a simulation study of the N191110 experiment.
\end{abstract}
\maketitle
\section{Introduction}
The inertial confinement fusion (ICF) is a primary experimental approach to study 
the high energy density fusion physics \cite{lan2022dream,chen2022determination}.
In 2022, a breakthrough has been made to produce 3.15 megajoules of fusion energy output from 
2.05 megajoules of energy to the target capsule and achieves the ICF ignition.
Following the Lawson criterion, 
an ignited plasma is one where the fusion self-heating power is high enough 
to overcome all the physical processes that cool the fusion plasma, 
creating a positive thermodynamic feedback loop with rapidly increasing temperature \cite{abu2022lawson}.
The plasma self-heating through the $\alpha$ particle transport and energy deposition
directly determines the fusion energy release and the implosion performance of the ICF capsule.
Therefore, an efficient and robust algorithm for the $\alpha$ particle transport
is highly required for the numerical simulation of ICF.

The $\alpha$ particle transport is a multi-scale process. 
On the scale of particle interaction, the $\alpha$ particle interacts with the 
background electron, deuterium, and tritium through the Coulomb potential.
On the mesoscopic scale, 
the evolution of the distribution function of charged particles follows the Boltzmann equation.
On the macroscopic scale, 
the $\alpha$ transport regime goes into the continuum regime, 
and a typical asymptotic analysis shows that the $\alpha$ transport process
can be approximated by the multi-group diffusion equations.
The particle number density and Coulomb cross-section are large in an ICF capsule,
which implies a high collision rate.
The effective Knudsen number, which is defined as the ratio of 
the effective $\pi/2$-collision mean free path to the characteristic length 
is large \cite{zylstra2019alpha}.
The velocity distribution of $\alpha$ particle is at a highly non-equilibrium state.
The kinetic Boltzmann equation with Coulomb potential needs to be solved.

To overcome the high dimension of the Boltzmann equation, we use the Monte Carlo method.
Although the effective free path of $\alpha$ particle is larger, 
the physical mean free path is much smaller than the ICF characteristic scale.
Therefore, the computation cost of the traditional Monte Carlo (MC)  method is 
extremely high and not applicable to the 2D/3D ICF simulation.
Since the 1950s, statistical models have been developing to predict 
the change in $\alpha$ particle velocity after multiple collisions \cite{bethe1953moliere,liu2017unified,liu2021unified}.
The statistic models work well for small-angle soft collisions, 
but the accuracy decreases when large-angle collisions take charge.
The energy deposition model has also been developed, 
including the effects of the short-range collision, 
the long-range interaction, 
and the quantum effect \cite{atzeni2004physics,brown2005charged}.
In this work, we propose a new hybrid collision model and a machine-learning energy deposition model
to improve the efficiency and accuracy of the $\alpha$ particle transport simulation.

The rest of this paper is organized as follows.
The physical model and kinetic equations of 
the $\alpha$ particle transport are introduced in Section \ref{section_model}.
In Section \ref{section_hybrid}, 
we introduce the hybrid collision model for $\alpha$ particle transport.
The machine learning based energy deposition model is presented in Section \ref{section_network}.
The numerical tests and ICF applications are shown in Section \ref{section_numerical}, 
and Section \ref{section_conclusion} is the conclusion.

\section{Physical model of charged particles transport}\label{section_model}
The evolution of $\alpha$ particle transport is characterized by 
the velocity distribution function $f(\vec{x},t,E,\vec{\Omega})$, 
where $\vec{x}$ is the spatial variable, 
$t$ is the time variable, 
$E$ is the particle energy, and 
$\vec{\Omega}$ is the velocity direction.
The velocity magnitude is related to the particle energy by $|\vec{v}|=\sqrt{2E/m_\alpha}$,
and $m_\alpha$ is the particle mass of $\alpha$ particle.
The transport process of $\alpha$ particle is described by the Boltzmann equation,
\begin{equation}\label{eq_Boltzamnn}
    \frac{\partial}{\partial t} f(\vec{x},t,E,\vec{\Omega}) +
    |\vec{v}|\vec{\Omega}\cdot\nabla f(\vec{x},t,E,\vec{\Omega}) =
    Q^+\left(f,f_\beta\right)+Q^-\left(f,f_\beta\right)+S,
\end{equation}
where $S$ is the $\alpha$ source from the local nuclear fusion
\begin{equation}\label{eq_nuclear}
    \begin{aligned}
        &D+T\to He^{4}(3.5 \text{ MeV})+n(14.1 \text{ MeV}),\\
        &D+He^3\to He^{4}(3.6 \text{ MeV})+p(14.7 \text{ MeV}).
    \end{aligned}
\end{equation}
In the nuclear fusion, the distribution of the $\alpha$ particle $He^4$ is 
\begin{equation}\label{eq_delta}
    f(\vec{x},t,E,\vec{\Omega})=\delta_{E_0}(E)U_{\mathcal{S}^2}(\vec{\Omega})N_\alpha(\vec{x},t),
\end{equation}
where $\delta_{E_0}(E)$ is the Dirac's delta function with $E_0=3.5 \text{MeV}$, 
$U_{\mathcal{S}^2}$ is the uniform distribution on the unit sphere $\mathcal{S}^2$, and 
$N_\alpha(\vec{x},t)$ is the generated number of $\alpha$ particle by the nuclear fusion. 
The interaction between the $\alpha$ particle and the background electron $e$, deuterium $D$, 
tritium $T$ is through the Coulomb interaction. 
The gain term $Q^+\left(f,f_\beta\right)$ for $\beta=e,D,T$ is
\begin{equation}\label{eq_gain}
    Q^+(f,f_\beta)=\int_{\mathcal{R}^3}\int_{\mathcal{S}^2}
    \sigma(|\vec{v}_r|,\vec{\Omega})|\vec{v}^{*}_r| 
    f_\beta(\vec{v}^{*}{}') f(\vec{v}^{*}) d\vec{v}^{*}d\vec{\Omega},
\end{equation}
and the loss term $Q^-\left(f,f_\beta\right)$
\begin{equation}\label{eq_loss}
    Q^-(f,f_\beta)=\int_{\mathcal{R}^3}\int_{\mathcal{S}^2}
    \sigma(|\vec{v}_r|,\vec{\Omega})|\vec{v}_r| 
    f_\beta(\vec{v}') f(\vec{v}) d\vec{v}'d\vec{\Omega}.
\end{equation}
The velocity pairs $(\vec{v}^{*}{}',\vec{v}^{*})$ and $(\vec{v}',\vec{v})$ are the reversible collision pairs in an elastic collision and $\vec{v}_r=\vec{v}-\vec{v}'$ is the relative velocity.
The differential cross-section of the Coulomb interaction is
\begin{equation}\label{eq_crosssectiond}
    d{\sigma_\beta}=\frac{b_{0}^{2}}{4}\frac{1}{{{\sin }^{4}}(\theta /2)}\sin \theta \mathrm{d}\theta \mathrm{d}\varphi,
\end{equation}
where $\theta$ is the collision polar angle and $\varphi$ is the collision azimuth angle.
The $\pi/2$ aiming distance $b_0$ is 
\begin{equation}\label{eq_b0}
    {{b}_{0}}=\frac{{{q}_{\alpha }}{{q}_{\beta }}}{4\pi {{\varepsilon }_{0}}\mu {{\vec{v}_r}^{2}}},
\end{equation}
where $\mu$ is the reduced mass
\begin{equation}\label{eq_mu}
    \mu=\frac{m_\alpha m_\beta}{m_\alpha+m_\beta}.
\end{equation}
The total collision cross-section is
\begin{equation}\label{eq_crosssectiont}
    {\Sigma }_{t}=\sum\limits_{\beta }{{{n}_{\beta }}\int_{0}^{2\pi }{\int_{{{\theta }_{p}}}^{\pi }{\frac{b_{0}^{2}}{4}\frac{1}{{{\sin }^{4}}(\theta /2)}\sin \theta d\theta d\varphi }}},
\end{equation}
where ${\theta }_{p}$ is the physical minimum scattering angle due to the Debye shielding, 
\begin{equation}
    {\theta }_{p}=\left\{
    \begin{aligned}
        &\sqrt{\hbar/\mu|\vec{v}_r|\lambda_D}\quad 
        &u/c\ge q_\alpha q_\beta / 2\pi\varepsilon_0\hbar c,\\
        &\sqrt{b_0/\lambda_D}\quad
        &u/c<q_\alpha q_\beta / 2\pi\varepsilon_0\hbar c,
    \end{aligned}
    \right.
\end{equation}
The Debye length $\lambda_D$ is 
\begin{equation}
    \lambda_D=\sqrt{\varepsilon_0T_e/n_e q_e^2},
\end{equation}
where $\varepsilon$ is the permittivity of free space, 
$n_e$ is the electron number density, 
$q_e$ is the charge of an electron, and
$T_e$ is the electron temperature.
The averaged streaming length of $\alpha$ transport between two successive collisions
is the mean free path $\lambda$, which can be calculated by
\begin{equation}\label{eq_mfp}
    \lambda=-\ln(\xi)/\Sigma_t,
\end{equation}
where $\xi$ is the random number uniformly distributed in $[0,1]$.
In the inertial confinement fusion, 
the plasma number density is $n_e\sim10^{3} \text{ nm}^{-3}$, 
the plasma temperature is $T_e\sim10^{2} \text{ MK}$,
the Debye length is $\lambda_D\sim 1 \text{ nm}$,
the $\pi/2$ aiming distance is $b_0\sim10^{-6} \text{ nm}$
the minimum scattering angle is $\theta_p\sim 10^{-4}$,
the total cross section $\sigma_t\sim10^{-3} \text{ nm}^{2}$.
The characteristic time of ICF is $t\sim10^{-1} \text{ ns}$ and 
the number of collisions of one $\alpha$ particle 
within the ICF characteristic time is $N_c> 10^{8}$.
Such a high collision rate of Coulomb interaction makes the numerical simulation extremely expensive. 

In a Coulomb collision between the high energy $\alpha$ particle and 
the background deuterium, tritium, electron, 
the $\alpha$ particle velocity direction $\Omega$ changes and 
a proportion of $\alpha$ particle energy will be deposited.
The change of velocity direction is characterized by 
the Coulomb collision cross section Eq.\eqref{eq_crosssectiond}, 
and simulated by the Monte Carlo method.
The Monte Carlo (MC) method is efficient for the high-dimensional 
calculation compared to the discrete ordinate method,
However, it suffers the stochastic noise with is on the order of $O(N_{MC}^{-0.5})$.
The $\alpha$ particle energy deposition is critical to the performance of ICF implosion, 
especially for the ICF designs at ignition cliff \cite{zylstra2022burning,craxton2015direct}.
Therefore, a precise analytical method is used to avoid 
the MC noise in the calculation of the energy deposition \cite{brown2005charged}.
In the following two sections, 
we present a hybrid collision model for the particle trajectory calculation 
and a neutral network for the energy deposition calculation.

\section{A hybrid collision model for high energy \texorpdfstring{$\alpha$}{} particle transport}
\label{section_hybrid}
\begin{figure}
  \centering
  \includegraphics[width=0.9\textwidth]{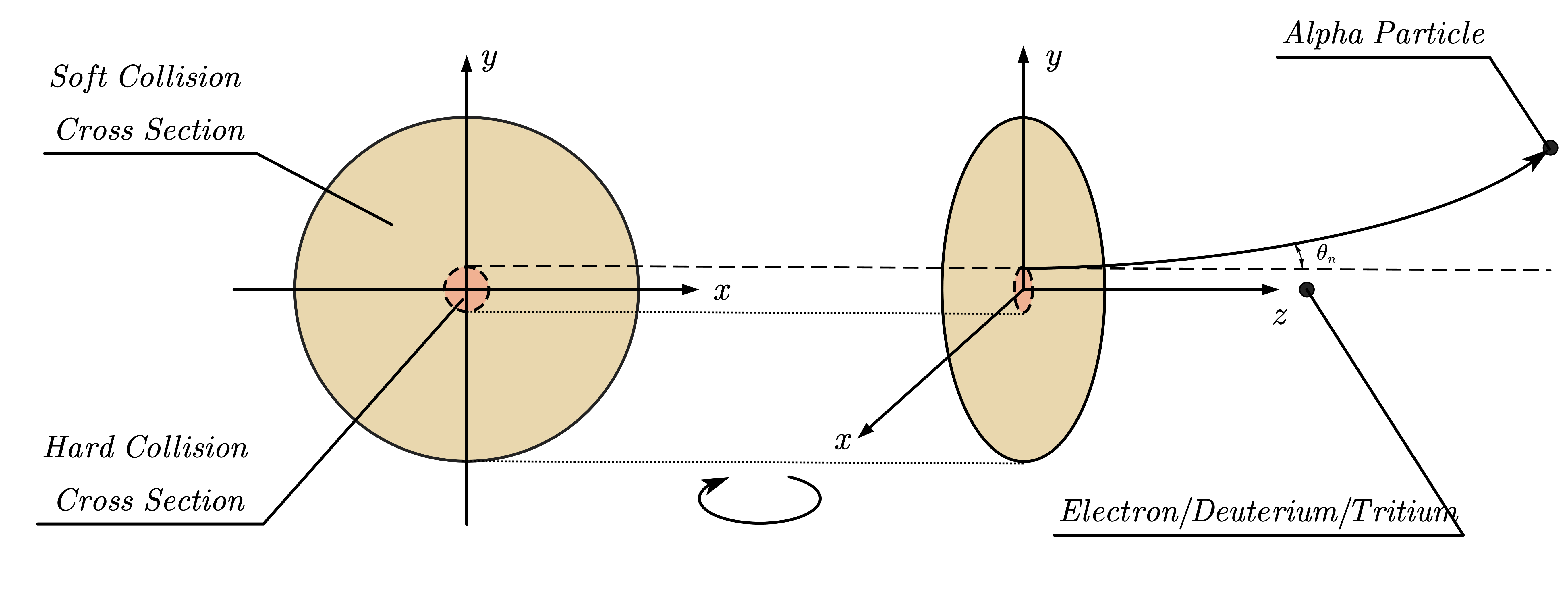}
  \caption{Hard collision cross-section, Soft collision cross-section, and numerical minimum scattering angle.}
  \label{fig_1}
\end{figure}
In this section, we introduce a hybrid collision model for $\alpha$ trajectory calculation 
to overcome the large computational cost of the high Coulomb-collision rate.
For Coulomb interaction, most collisions are grazing collisions with a small scattering polar angle,
and only a small proportion of collisions are hard collisions whose scattering polar angle is large.
For soft collisions, the aiming distance of the $\alpha$ particle and the target deuterium/tritium/electron particle is relatively large, and for hard collisions, the aiming distance is small.
We introduce an artificial scattering-angle parameter $\theta_n$ and 
correspondingly, an aiming-distance parameter $b_n$,
to characterize the collision type and the cross-section.
Define the collisions with a scattering angle smaller than $\theta_n$ as the soft collision,
and define the collision with a scattering angle larger than $\theta_n$ as the hard collision.
In the simulation of inertial confinement fusion, 
the parameter of the numerical minimum scattering angle is defined as 
\begin{equation}\label{eq_thetan}
    \theta_n=\arccot\left(N_h
    \frac{4\varepsilon\mu }{q_\alpha q_e}
    \sqrt{\frac{\pi v_\alpha}{\Delta t n_e}}
    \right),
\end{equation}
where $N_h$ is the number of hard collisions in a time step $\Delta t$.
Typically, the scattering-angle parameter is chosen as $\theta_n\sim10^{-3}$, and
the corresponding aiming-distance parameter is $b_{0,n}\sim 1\text{ nm}$.
For a time step $\Delta t\sim 10^{-3} \text{ ns}$,
the hard collision number is estimated to be $N_h\sim10^{2}$ and 
the soft collision number is estimated to be $N_s\sim10^{5}$.
For hard collisions, the typical Monte Carlo method is used.
Based on the numerical minimum scattering angle ${{\theta }_{n}}$ 
the collision cross-section is divided into 
the hard scattering angle cross section $\theta \le {{\theta }_{n}}$ and 
the soft scattering angle cross-section 
${{\Sigma }_{t}}={{\Sigma }_{th}}+{{\Sigma }_{ts}}$ as shown in figure \ref{fig_1}.
The hard scattering cross section ${{\Sigma }_{th}}$ between $\alpha$ particle and $\beta=e/D/T$ particle is
\begin{equation}\label{eq_crosssectionh}
    \Sigma_{th}=n_\beta \int_0^{2\pi} \int_{\theta_p}^{\theta_N} \frac{b_0^2}{4}
    \frac{1}{\sin^4(\theta/2)}\sin \theta \mathrm{d} \theta \mathrm{d} \varphi, 
\end{equation}
and the soft scattering cross section ${{\Sigma }_{ts}}$ is
\begin{equation}\label{eq_crosssections}
    \Sigma_{ts}=n_\beta \int_0^{2\pi} \int_{\theta_N}^{\pi} \frac{b_0^2}{4}
    \frac{1}{\sin^4(\theta/2)}\sin \theta \mathrm{d} \theta \mathrm{d} \varphi.
\end{equation}
The particle free path and particle collision time of hard collisions are
calculated by the hard collision cross section Eq.\eqref{eq_crosssectionh},
and the scattering vector is sampled from the Coulomb cross-section equation \ref{eq_crosssectiond}.
The numerical minimum scattering angle Eq.\eqref{eq_thetan} is chosen such that $N_h$ hard collisions happen in a time step $\Delta t$.
Between two hard collisions,
an efficient statistical model is developed to calculate the numerous soft collisions,
and the statistical model is stated as the following theorem.
\begin{spacing}{1.5}
\begin{theorem}
    Define the time interval between two hard collisions as the hard collision free time. 
    Consider a hard collision free time $\tau_h$, in which $N_s$ successive soft collisions happen 
    between $\alpha$ particle and background $\beta$ particle, for $\beta=D,T,e$. 
    Assume the energy of $\alpha$ particle is high with $|\vec{v}_\alpha|\gg|\vec{v}_\beta|$.
    The $\alpha$ particle velocity direction vectors are $\vec{\Omega}_i$, for $i=1,...,N_s$.
    Define the scattering vector $\vec{T}_{i}$ as 
    $\vec{T}_{i}=\vec{\Omega}_{i+1}-\vec{\Omega}_i$, for $i=1,...,N_s-1$.
    We have \begin{description}
    \item [(i)]  
    \begin{equation}\label{eq_softtotal}
        \vec{\Omega}_{N_s}=\vec{\Omega}_{1}+N_s \vec{T}_{1}+O(\Delta t^{5/2});
    \end{equation}
    \item [(ii)] The polar scattering angle $\Theta_{N_s}$ between $\vec{\Omega}_{N_s}$ and 
    $\vec{\Omega}_{1}$ follows a Gaussian distribution
    \begin{equation}\label{eq_gaussian}
        f(\Theta)=\frac{1}{\sqrt{2\pi\left<\Theta^2\right>}}\exp\left(-\frac{\Theta^2}{2\left<\Theta^2\right>}\right)+O(N_s^{-2/5})+O(\Delta t^{3/2}),
    \end{equation}
    with a variance
    \begin{equation}
    \left<\Theta^2\right>=4\pi n_\beta |\vec{v}_\alpha|\tau_h b_0^2 \left.\ln\left[\sin\left(\frac{\theta}{2}\right)\right]\right|_{\theta_p}^{\theta_n}.
    \end{equation}
    The azimuth angle $\Phi_{N_s}$ between $\vec{\Omega}_{N_s}$ and 
    $\vec{\Omega}_{1}$ follows a uniform distribution in $[0,2\pi]$
    \begin{equation}\label{eq_uniform}
        f(\Phi)=\frac{1}{2\pi}.
    \end{equation}
    \end{description}
\end{theorem}
\begin{figure}
  \centering
  \includegraphics[width=0.3\textwidth]{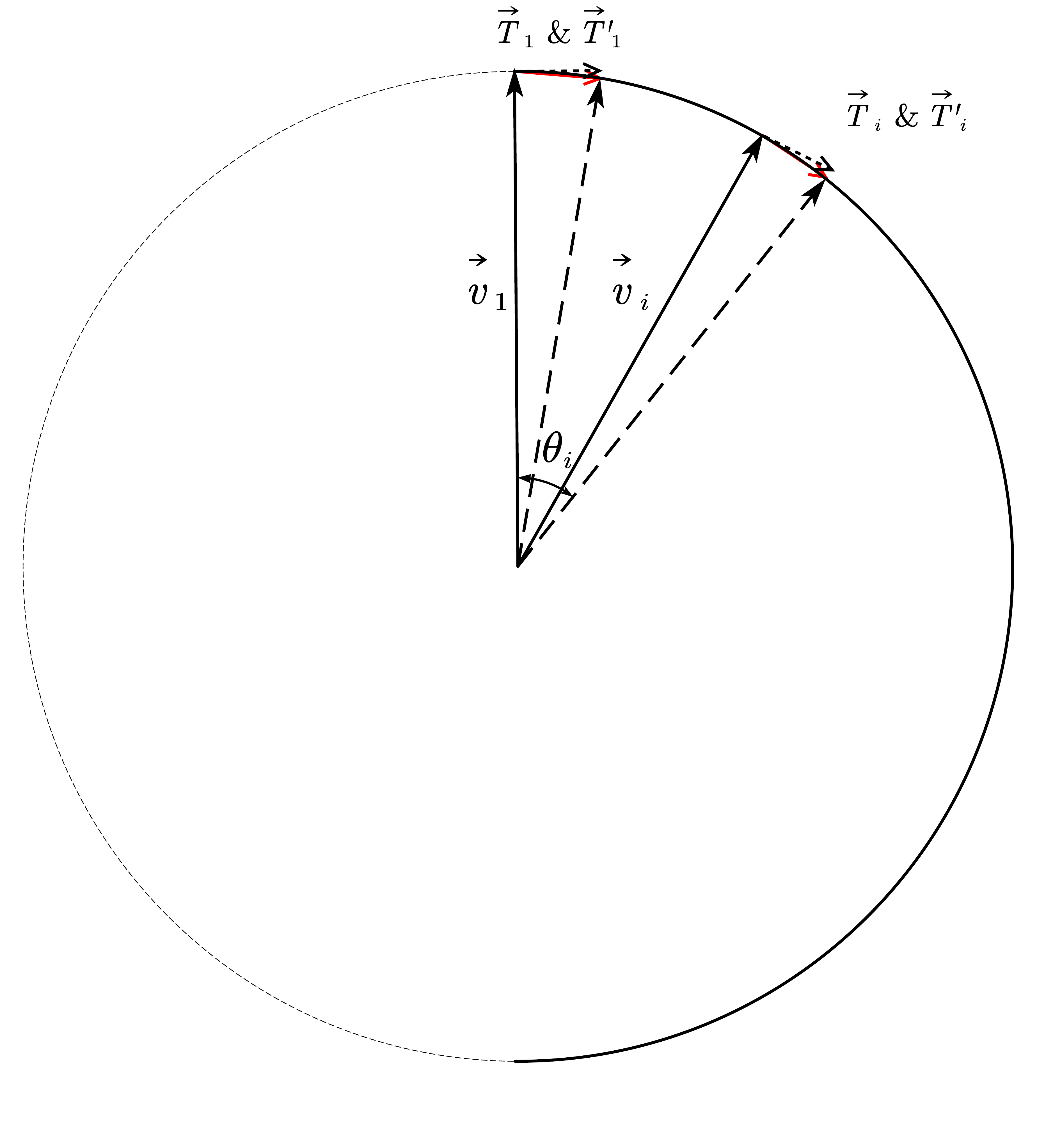}
  \caption{The polar angle scattering in the soft collision processes.}
  \label{fig_crosssection}
\end{figure}
\begin{proof}
    Consider the $\alpha$ particle with energy much higher than the background particles,
    i.e., $|\vec{v}_\alpha|\gg|\vec{v}_\beta|$.
    We have $\vec{v}_\alpha\sim \vec{v}_r$ where $\vec{v}_r$ is the relative velocity.
    Therefore, in the collision process, 
    the polar and azimuth scattering angles of $\vec{v}_\alpha$ 
    follow the Coulomb cross section Eq.\eqref{eq_crosssectiond}.
    
    (i) The scattering vector $\vec{T}_i$ transfers $\vec{\Omega}_i$ to $\vec{\Omega}_{i+1}$.
    Define an adjoint $\vec{T}'_i$ such that $\vec{T}'_i$ is perpendicular to $\vec{\Omega}_i$, 
    and $|\vec{T}'_i|=|\vec{T}_i|$, as shown in figure \ref{fig_crosssection}.
    The polar scattering angle between $\vec{\Omega}_i$ and $\vec{\Omega}_{i+1}$ is $\theta_n$.
    It can be calculated that the angle between $\vec{T}_i$ and $\vec{T}'_i$ is $\theta_n/2$, 
    and 
    \begin{equation}
        |\vec{T}'_i-\vec{T}_i|=\frac{1}{2}\theta_n|\vec{T}_i|=O(\theta_n).
    \end{equation}
    According to equation \eqref{eq_thetan} $\theta_n\sim(\Delta t^{1/2})$, 
    and therefore $|\vec{T}'_i-\vec{T}_i|\sim O(\Delta t^{1/2})$.
    The angle between $\vec{T}'_{N_s}$ and $\vec{T}'_1$ is $N_s\theta_n$, and
    the difference between $\vec{T}'_{N_s}$ and $\vec{T}'_1$ is calculated as
    \begin{equation}
        |\vec{T}'_{N_s}-\vec{T}'_1|<|\vec{T}_1|N_s\theta_n=O(N_s\theta_n).
    \end{equation}
    The total number of soft collision is $N_s \sim O(\Sigma_{ts}|\vec{v}_\alpha| \tau_h)$,
    where $\Sigma_{ts}$ is the total soft collision cross section.
    Based on equation \eqref{eq_thetan}, the soft collision cross section Eq.\eqref{eq_crosssections} can be expanded as $\Sigma_{ts}\sim O(\Delta t)$.
    Therefore $N_s\sim O(\Delta t)$ and $|\vec{T}'_{N_s}-\vec{T}'_1|\sim{\Delta t^{3/2}}$.
    The difference between $\vec{T}_i$ and $\vec{T}_1$ can be estimated as
    \begin{equation}
        \begin{aligned}
            |\vec{T}_{N_s}-\vec{T}_1|<&|\vec{T}_{N_s}-\vec{T}'_i|+
            |\vec{T}'_{N_s}-\vec{T}'_1|+|\vec{T}'_{N_s}-\vec{T}_1|\\
            \sim &O((N_s+2)\theta_n)
            \sim O(\Delta t^{3/2}).
        \end{aligned}
    \end{equation}
    The total velocity directional change can be calculated as
    \begin{equation}
        \begin{aligned}
        \vec{\Omega}_{N_s}=&\vec{\Omega}_{1}+\sum_{i=1}^{N_s} \vec{T}_{i}\\
            <&\vec{\Omega}_{1}+N_s\vec{T}_{1} +N_s |\vec{T}_{N_s}-\vec{T}_1|\\
            = &\vec{\Omega}_{1}+N_s\vec{T}_{1} +O(\Delta t^{5/2})
        \end{aligned}
    \end{equation}
    Physically, it is estimated that the total scattering angle $\theta_T$ of $\alpha$ particle is small 
    $\theta_T<10^{-2}$ until it decreases to the local plasma thermal velocity \cite{atzeni2004physics},
    therefore the absolute error in equation \eqref{eq_softtotal} is acceptable.
    
    The scattering vector of a soft collision can be decomposed as 
    a polar scattering vector and an azimuthal scattering vector, namely
    \begin{equation}
        \vec{T}_i=\vec{\theta}_i+\vec{\varphi}_i, \quad \text{for} \quad i=1,...,N_s.
    \end{equation}
    The polar scattering and azimuth scattering are commutative, i.e.,
    \begin{equation}
        \begin{aligned}
            \vec{\Omega}_{N_s}=&\vec{\Omega}_1+\sum_{i=1}^{N_s}(\vec{\theta}_i+\vec{\varphi}_i)+O(\Delta t^{5/2})\\
            =&\vec{\Omega}_1+\sum_{i=1}^{N_s}\vec{\theta}_i+\sum_{i=1}^{N_s}\vec{\varphi}_i+O(\Delta t^{5/2}),
        \end{aligned}
    \end{equation}
    where $\Theta_{N_s}=\sum_{i=1}^{N_s}\vec{\theta}_i$ and $\Phi_{N_s}=\sum_{i=1}^{N_s}\vec{\varphi}_i$.
    The probability density distribution of $\vec{\theta}_i$ is 
    \begin{equation}
        f(\theta_i)=\frac{1}{2[\csc^2(\pi/2)-\csc^2(\theta_n/2)]}
        \frac{\sin(\theta_i)}{\sin^4(\theta_i/2)}
    \end{equation}
    According to the central limit theorem, the probability density distribution of 
    $\Theta_{N_s}=\sum_{i=1}^{N_s}\vec{\theta}_i$ follows
    \begin{equation}\label{eq_distributiontheta}
        f(\Theta)=\frac{1}{\sqrt{2\pi\left<\Theta^2\right>}}\exp\left(-\frac{\Theta^2}{2\left<\Theta^2\right>}\right)+O(N_s^{-2/5})+O(\Delta t^{3/2}),
    \end{equation}
    and the variance is calculated following Moliere's multiple scattering theory \cite{bethe1953moliere,particle2022review}
    \begin{equation}\label{eq_variance}
        \begin{aligned}
            \left\langle {{\Theta }^{2}} \right\rangle
            =&2 n_\beta |\vec{v}_\alpha|\tau_h
            \underbrace{\frac{\mu^2}{m_\alpha^2}\int_{0}^{2\pi}
            \int_{\theta_p}^{\theta_n}
            (1-\cos(\theta))\sigma(\theta)\sin(\theta) \mathrm{d}\theta 
            \mathrm{d}\varphi}_{\text{momentum transfer cross-section}} \\
            =&4\pi n_\beta |\vec{v}_\alpha|\tau_h b_0^2 \left.\ln\left[\sin\left(\frac{\theta}{2}\right)\right]\right|_{\theta_p}^{\theta_n}.
    \end{aligned}
    \end{equation}
    Here, the momentum transfer cross-section calculates the momentum change of the $\alpha$ particle in the direction of its initial motion.
    Eq.\ref{eq_variance} calculates the momentum change in the initial direction in $\tau_h$, which equals the second order moment of $\Theta$.
    The probability density distribution of the polar scattering angle $\Theta_{N_s}$ between $\vec{\Omega}_{N_s}$ and $\vec{\Omega}_{1}$ are given in equations \eqref{eq_distributiontheta} 
    and \eqref{eq_variance}.
    
    The probability density distribution of $\vec{\varphi}_i$ is
    \begin{equation}
        f(\varphi_i)=\frac{1}{2\pi}.
    \end{equation}
    The azimuth angle $\Phi_{N_s}=\sum_{i=1}^{N_s}\vec{\varphi}_i$ can be written as
    \begin{equation}
        \Phi_{N_s}=\sum_{i=1}^{N_s}{\varphi_i}-2\pi\lfloor\sum_{i=1}^{N_s}{\varphi_i}/2\pi\rfloor.
    \end{equation}
    We use the mathematical induction to derive the distribution of $\Phi_{N_s}$.
    Assuming that  $\Phi_{k}$ uniformly distributes in $[0,2\pi]$, then we have a
    a one-to-one mapping from $\Phi_{k+1}$ to $\varphi_k$,
    \begin{equation}
        \Phi_{k+1}=\Phi_{k}+\varphi_k-2\pi\lfloor(\Phi_{k}+\varphi_k)/2\pi\rfloor\in[0,2\pi],
    \end{equation}
    which indicates that $\Phi_{k+1}$ follows the same distribution with $\varphi_k$.
    Therefore, the azimuth angle $\Phi_{N_s}$ between $\vec{\Omega}_{N_s}$ and 
    $\vec{\Omega}_{1}$ uniform distributes in $[0,2\pi]$.
\end{proof}
\end{spacing}

\begin{figure}
  \centering
  \includegraphics[width=0.6\textwidth]{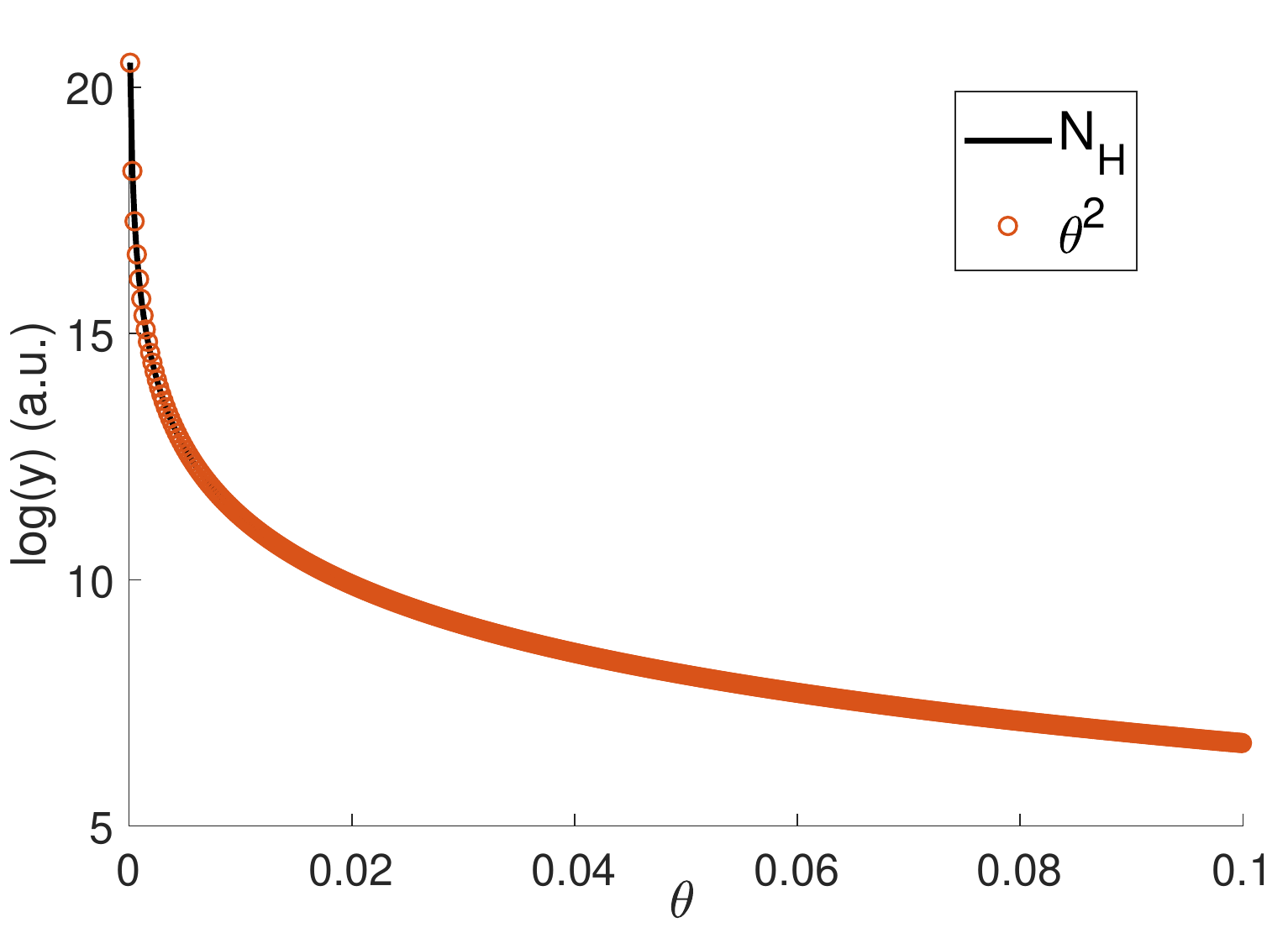}
  \caption{The number of hard collisions decreases as $N_h\sim \theta^{-2}$.}
  \label{fig_crosstheta}
\end{figure}
In a time step $\Delta t$, the number of hard collisions is calculated as
\begin{equation}
    N_h(\theta_N) =\Delta t |\vec{v}_\alpha|n_\beta\int_0^{2\pi}\int_{\theta_N}^{\pi}\frac{b_0^2}{4}\frac{1}{\sin^4(\theta/2)}\sin(\theta)\mathrm{d}\theta\mathrm{d}\varphi.
\end{equation}
We interpolate the $N_h(\theta_N) $ for $\theta\in[10^{-4},10^{-1}]$, which gives
\begin{equation}
    N_h(\theta_N) \sim \frac{\Delta t |\vec{v}_\alpha|n_\beta b_0^2}{2}\theta^{-2},
\end{equation}
as shown in figure \ref{fig_crosstheta}.
The Monte Carlo computational cost of the hybrid method is proportional to the hard collision rate, 
and therefore decreases as $\theta^{-2}$.
The algorithm for the $\alpha$ particle velocity direction scattering is shown in algorithm 1.
\begin{algorithm}[H]\label{algorithm1}
    \caption{Algorithm for the $\alpha$ particle velocity direction scattering}
    \begin{algorithmic}[1]
        \FOR {$ t \in [t^n,t^{n+1}]$ }
        \STATE {Sample hard collision free time $\tau_h$ for $\alpha$ particle transport in $e/D/T$ material;}
        \STATE {Calculate the hard collision by the Coulomb cross section equation \ref{eq_crosssectiond};}
        \STATE {Calculate the soft collisions by the statistic model equations \ref{eq_gaussian} and \ref{eq_uniform};}
        \STATE {Advance time by the hard collision free time $ t= t +\tau_h $.}
        \ENDFOR
    \end{algorithmic}
\end{algorithm}

\subsection{Neural network model for the \texorpdfstring{$\alpha$}{} particle energy deposition}
\label{section_network}
\begin{figure}
  \centering
  \includegraphics[width=0.65\textwidth]{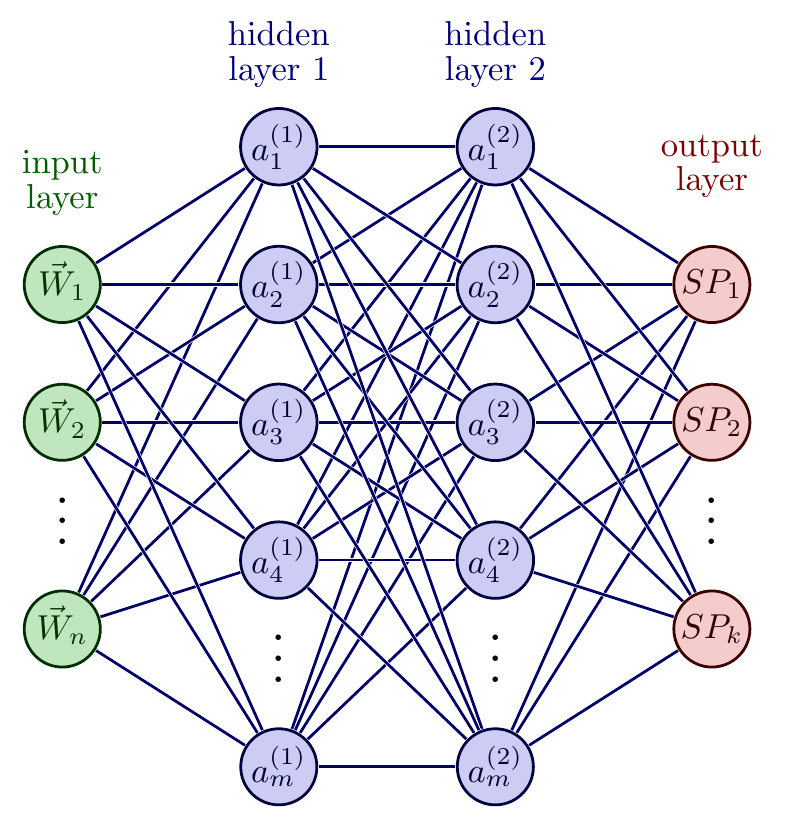}
  \caption{The structure of the stopping power neural network.}
  \label{fig_neutralnet}
\end{figure}

\begin{figure}
  \centering
  \includegraphics[width=0.45\textwidth]{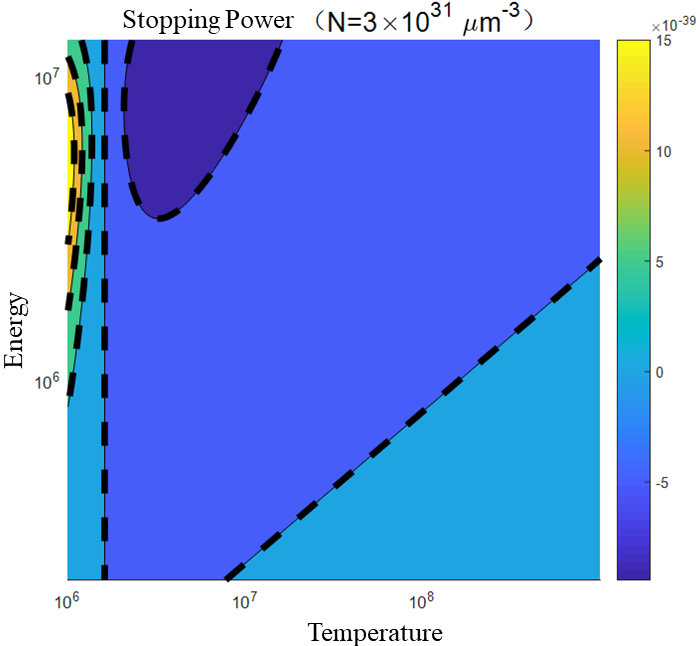}
  \includegraphics[width=0.5\textwidth]{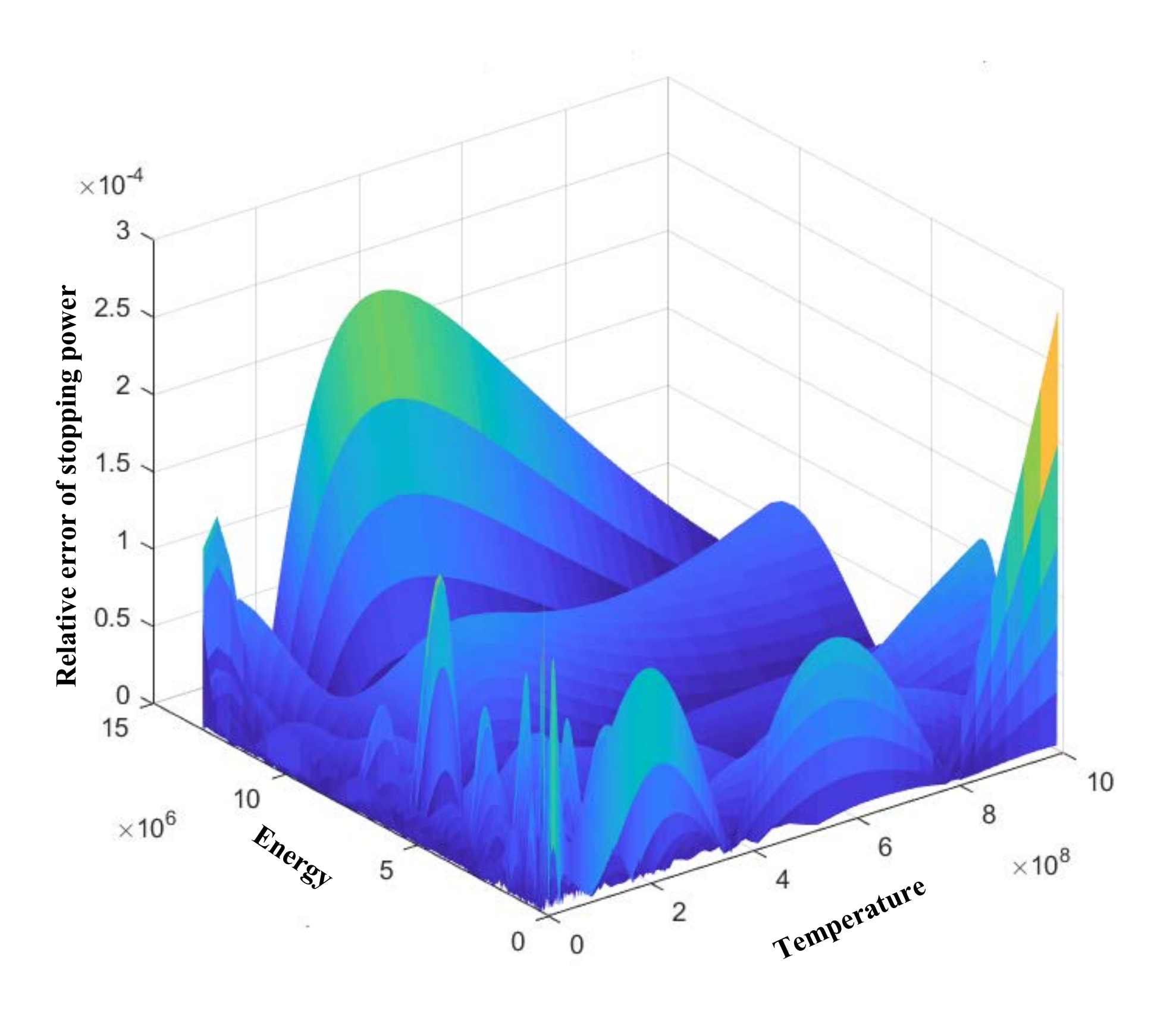}
  \caption{The comparison between the neutral network interpolation and BPS data is shown in the left figure; the contours are the BPS data, and the dashed lines are the neutral network interpolation. The relative error is shown in the right figure.}
  \label{fig_train}
\end{figure}

In the ICF process, the $\alpha$ particle deposits its energy to 
the hot spot and D/T ice shell as it transports in the ICF capsule. 
In the high energy state, 
assuming that the $\alpha$ particle velocity is much larger than the local thermal velocity,
the $\alpha$ particle energy deposition rate can be derived from the Boltzmann equation.
The stopping power model can be written as \cite{brown2005charged}
\begin{equation}
    {{\left\langle \frac{d{{E}_{\alpha }}}{ds} \right\rangle }_{\beta }}=-\frac{{{n}_{\beta }}{{({{q}_{\alpha }}{{q}_{\beta }})}^{2}}}{4\pi \varepsilon _{0}^{2}{{m}_{\beta }}v_{\alpha }^{2}}\ln \overline{\Lambda }\left[\Phi ({{v}_{\alpha }}/{{v}_{\beta T}})-(1+\frac{{{m}_{\beta }}}{{{m}_{\alpha }}})\frac{2}{\sqrt{\pi }}\frac{{{v}_{\alpha }}}{{{v}_{\beta T}}}\exp (-{{(\frac{{{v}_{\alpha }}}{{{v}_{\beta T}}})}^{2}})\right],
\end{equation}
Where $\beta=e/D/T$ is the background material medium, 
$v_{\beta,T}$ is the thermal velocity of $\beta$ medium,
$\ln \overline{\Lambda }$ is the Coulomb logarithm, 
${{q}_{\alpha }},{{q}_{\beta }}$ is the amount of charge carried by the particle, 
${{\varepsilon }_{0}}$ is the vacuum permittivity, and $\mu $ is the reduced mass. 
The common Coulomb logarithm models are the Lee-More model, the Atzeni model, the Spitzer-Harm model, 
and the Brown-Preston-Singleton (BPS) model 
\cite{atzeni2004physics,brown2005charged}.
The Coulomb logarithm of the Lee-More model is
\begin{equation}
    \ln \overline{\Lambda }=\frac{1}{2}\ln (1+\frac{\lambda _{D}^{2}}{\left\langle b_{\min }^{2} \right\rangle }),
\end{equation}
where
\begin{equation}
    {{b}_{\min }}=\max (\frac{{{q}_{\alpha }}{{q}_{\beta }}}{4\pi {{\varepsilon }_{0}}\mu {{u}^{2}}},\ \frac{{{\lambda }_{dB}}}{4\pi })
\end{equation}
is the modified minimum aiming distance of collision, 
and $\lambda _{D}^{{}}$ is Debye length. 
The Coulomb logarithm of the Atzeni model is
\begin{equation}
    \begin{aligned}
  & \ln {{\overline{\Lambda }}_{i,e}}=7.1-0.5\ln {{n}_{e}}[{{10}^{21}}/c{{m}^{3}}]+\ln {{T}_{e}}[keV], \\ 
 & \ln {{\overline{\Lambda }}_{i,i}}=9.2-0.5\ln {{n}_{e}}[{{10}^{21}}/c{{m}^{3}}]+1.5\ln {{T}_{i}}[keV]. 
\end{aligned}
\end{equation}
The Coulomb logarithm of the Spitzer-Harm model is
\begin{equation}
    \ln \overline{\Lambda }=\ln \left\langle \frac{{{\lambda }_{D}}}{{{b}_{\min }}} \right\rangle \approx \ln \frac{{{\lambda }_{D}}}{\left\langle {{b}_{\min }} \right\rangle }.
\end{equation}
In the ICF plasma state, the BPS model reveals more detailed physics and the Coulomb logarithm is
\begin{equation}
    \frac{dE}{dx}=\frac{dE_{\beta ,S}^{C}}{dx}+\frac{dE_{\beta ,R}^{<}}{dx}+\frac{dE_{\beta }^{Q}}{dx},
\end{equation}
where the short-range collision term is
\begin{equation}
    \begin{aligned}
  & \frac{dE_{\beta ,S}^{C}}{dx}=\frac{e_{\alpha }^{2}}{4\pi {{\varepsilon }_{0}}}\frac{k_{\beta }^{2}}{{{m}_{\alpha }}{{v}_{\alpha }}}\sqrt{\frac{{{m}_{\beta }}}{2\pi {{\tau }_{\beta }}}}\int_{0}^{1}{du\sqrt{u}}\exp (-\frac{1}{2}{{\tau }_{\beta }}{{m}_{\beta }}v_{\alpha }^{2}u)\times  \\ 
 & \left\{ [-ln({{\tau }_{\beta }}\frac{\left| {{e}_{\alpha }}{{e}_{\beta }} \right|K}{4\pi {{\varepsilon }_{0}}}(\frac{{{m}_{\beta }}}{{{m}_{\alpha }}}+1)\frac{u}{1-u})+2-2\gamma ][({{m}_{\alpha }}+{{m}_{\beta }}){{\tau }_{\beta }}v_{\alpha }^{2}-\frac{1}{u}]+\frac{2}{u} \right\}.
\end{aligned}
\end{equation}
The long-range collision term is
\begin{equation}
    \begin{aligned}
  & \frac{dE_{\beta ,R}^{<}}{dx}=\frac{e_{\alpha }^{2}}{4\pi {{\varepsilon }_{0}}}\frac{i}{2\pi }\int_{-1}^{+1}{d\cos \theta \cos \theta \frac{{{\rho }_{\beta }}({{v}_{\alpha }}\cos \theta )}{{{\rho }_{total}}({{v}_{\alpha }}\cos \theta )}}F({{v}_{\alpha }}\cos \theta )ln(\frac{F({{v}_{\alpha }}\cos \theta )}{{{K}^{2}}}) \\ 
 & -\frac{e_{\alpha }^{2}}{4\pi {{\varepsilon }_{0}}}\frac{i}{2\pi }\frac{1}{{{\tau }_{\beta }}{{m}_{\alpha }}v_{\alpha }^{2}}\frac{{{\rho }_{\beta }}({{v}_{\alpha }})}{{{\rho }_{total}}({{v}_{\alpha }})}[F({{v}_{\alpha }})ln(\frac{F({{v}_{\alpha }})}{{{K}^{2}}})-{{F}^{*}}({{v}_{\alpha }})ln(\frac{{{F}^{*}}({{v}_{\alpha }})}{{{K}^{2}}})],
\end{aligned}
\end{equation}
where
\begin{equation}
    F(x)=-\int_{-\infty }^{\infty }{dv\frac{{{\rho }_{total}}(v)}{x-v+i\eta }},
\end{equation}
and $\rho_{total}=\rho_e+\rho_D+\rho_T$.
The quantum correction term is
\begin{equation}
    \begin{aligned}
  & \frac{dE_{\beta }^{Q}}{dx}=\frac{e_{\alpha }^{2}}{4\pi {{\varepsilon }_{0}}}\frac{k_{\beta }^{2}}{2{{\tau }_{\beta }}{{m}_{\beta }}v_{\alpha }^{2}}\sqrt{\frac{{{\tau }_{\beta }}{{m}_{\beta }}}{2\pi }}\int_{0}^{\infty }{d{{v}_{\alpha \beta }}}\times [2Re\psi (1+i{{\eta }_{\alpha \beta }})-ln\eta _{\alpha \beta }^{2}]\times  \\ 
 & \left\{ [1+(\frac{{{m}_{\alpha }}}{{{m}_{\beta }}}+1)\frac{{{v}_{\alpha }}}{{{v}_{\alpha \beta }}}(\frac{1}{{{\tau }_{\beta }}{{m}_{\beta }}{{v}_{\alpha }}{{v}_{\alpha \beta }}}-1)]\exp [-\frac{1}{2}{{\tau }_{\beta }}{{m}_{\beta }}{{({{v}_{\alpha }}-{{v}_{\alpha \beta }})}^{2}}] \right. \\ 
 & \left. -[1+(\frac{{{m}_{\alpha }}}{{{m}_{\beta }}}+1)\frac{{{v}_{\alpha }}}{{{v}_{\alpha \beta }}}(\frac{1}{{{\tau }_{\beta }}{{m}_{\beta }}{{v}_{\alpha }}{{v}_{\alpha \beta }}}+1)]\exp [-\frac{1}{2}{{\tau }_{\beta }}{{m}_{\beta }}{{({{v}_{\alpha }}+{{v}_{\alpha \beta }})}^{2}}] \right\}.
\end{aligned}
\end{equation}
The BPS model takes into account more detailed physics, 
however, the high dimension integral greatly limits its computational efficiency.
We use a fully connected neural network to interpolate the BPS stopping power model.
The neural network consists of two inner layers with ten neutrons in each layer.
The input layer contains the density, velocity, and temperature of electrons, deuterium, and tritium.
The tansig function is used as the activation function.
The structure of the neural network is shown in figure \ref{fig_neutralnet}.
The Lenvenberg-Marquardt method is used to train the stopping power neutral network.
The mean square error reaches $10^{-9}$ after about $1000$ epochs iteration.
The comparison between the neutral network interpolation and BPS data is shown in figure \ref{fig_train}.

\begin{figure}
  \centering
  \includegraphics[width=0.95\textwidth]{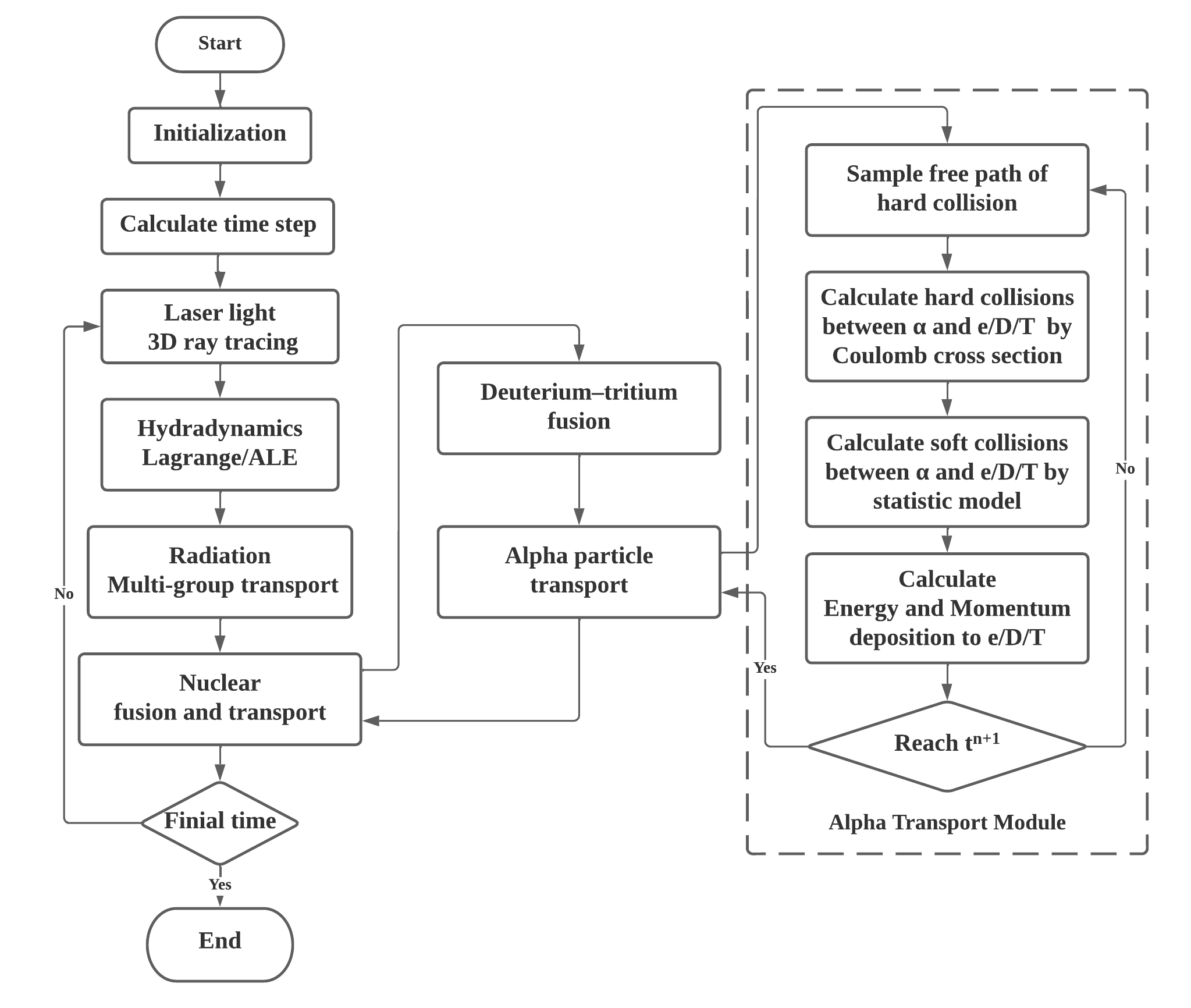}
  \caption{The flow chart of the $\alpha$ particle transport algorithm.}
  \label{fig_flowchart}
\end{figure}

The numerical method for $\alpha$ particle transport is composed of 
the hybrid collision model that describes the velocity scattering and direction change and 
the neural network predicts the energy deposition. 
The algorithm is shown in the flow chart \ref{fig_flowchart}.

\section{Numerical tests and applications in the ICF simulation}
\label{section_numerical}
\subsection{Numerical tests for the \texorpdfstring{$\alpha$}{} particle transport in a homogeneous medium}
\begin{figure}
  \centering
  \includegraphics[width=0.5\textwidth]{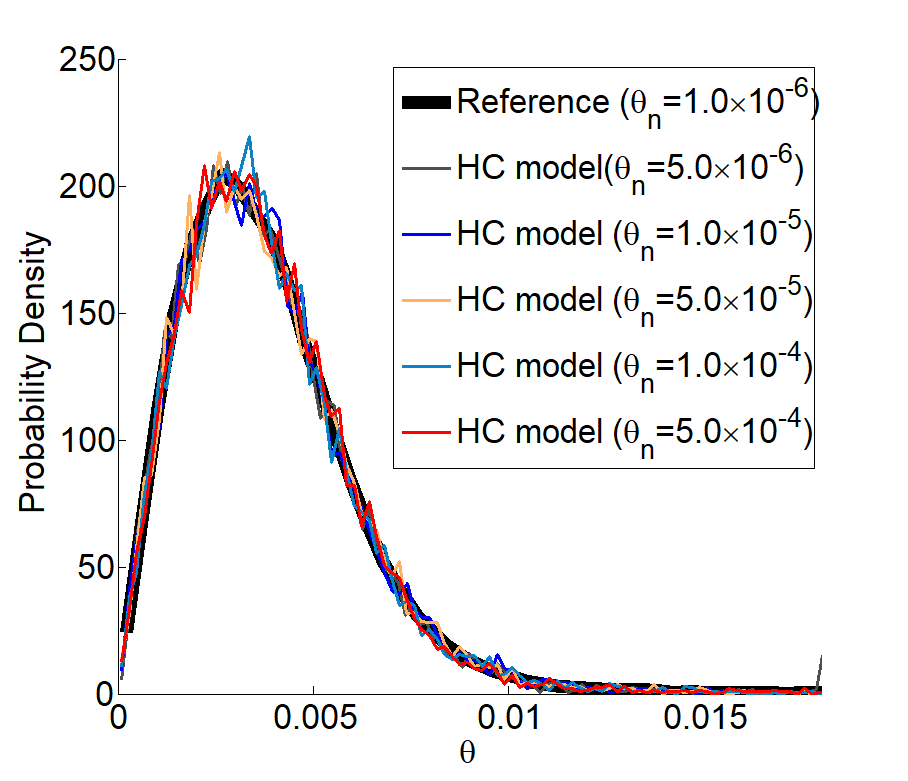}
  \includegraphics[width=0.45\textwidth]{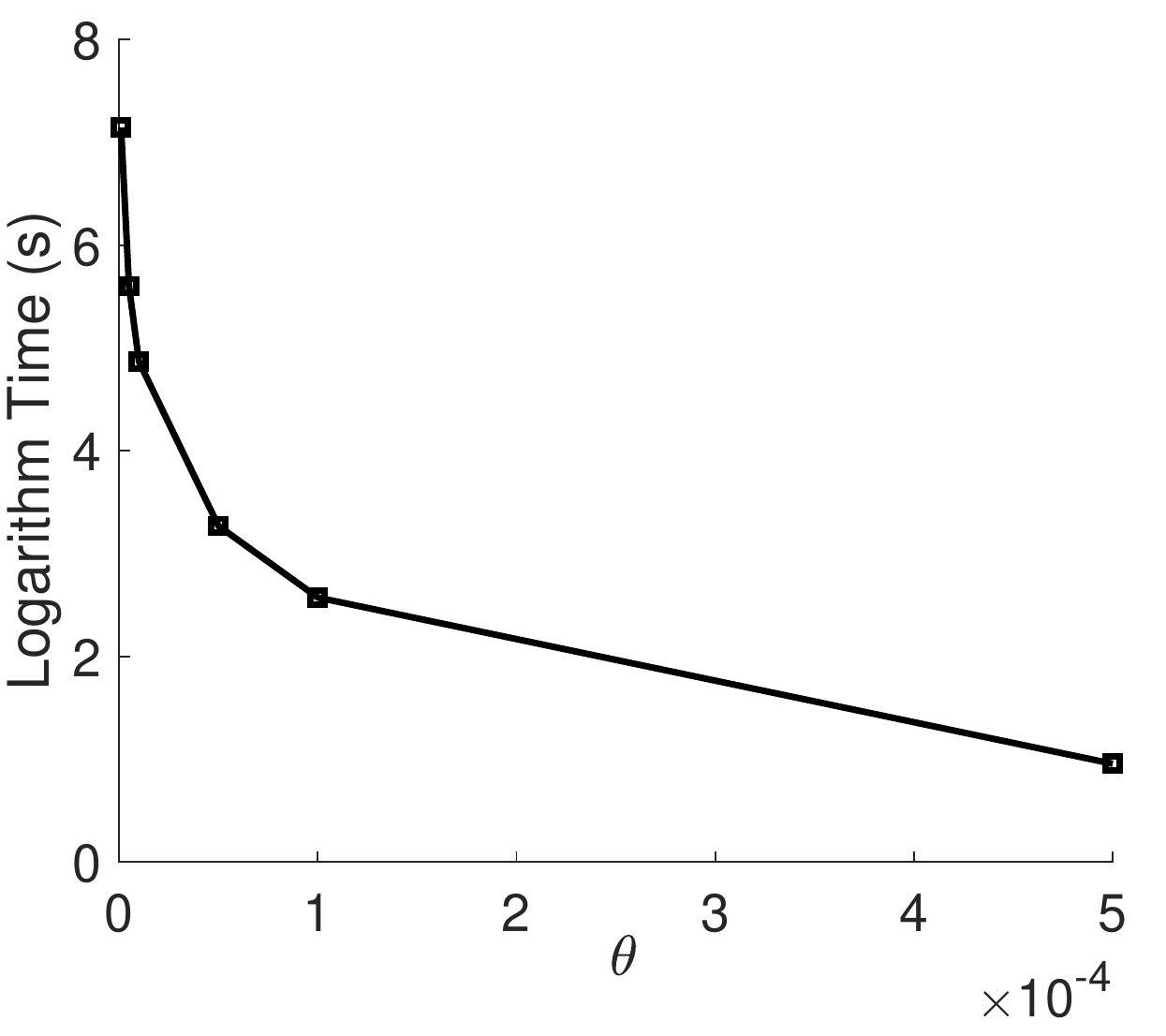}
  \caption{The comparison of the velocity distribution with different $\theta_n$ is shown in the left figure,
  and the computational time is shown in the right figure.}
  \label{fig_homo}
\end{figure}

To verify the accuracy and efficiency of the hybrid collision model, 
we simulate the transport process of a charged particle in a uniform background plasma. 
The initial distribution of the charged particle is $f(\vec{v})=\delta (1)$, 
the physical free path of the charged particle is $\lambda={{10}^{-8}}$, and 
the numerical minimum scattering angle is set as
${{\theta }_{n}}=1.0\times {{10}^{-6}},5.0\times {{10}^{-6}},1.0\times {{10}^{-5}},5.0\times {{10}^{-5}},1.0\times {{10}^{-4}},5.0\times {{10}^{-4}}$.
We simulate the $\alpha$ particle velocity distribution function for a transport distance of $x=1$.
The comparison of the velocity distribution with different $\theta_n$ is shown in figure \ref{fig_homo}, and 
the calculation time is shown in table \ref{table1}. 
It is shown that the hybrid collision model is not sensitive to the numerical parameter $\theta_n$, and 
the computational efficiency can be improved by two orders of magnitude.
\begin{table}
\centering
\begin{tabular}
{|c|c|c|c|c|c|c|} \hline
$\theta_n$& $5.0\times10^{-4}$ 
& $1.0\times10^{-4}$
& $5.0\times10^{-5}$
& $1.0\times10^{-5}$
& $5.0\times10^{-6}$
& $1.0\times10^{-6}$
\\ \hline
CPU time& $2.6s$ 
& $13.1s$
& $26.3s$
& $130s$
& $271s$
& $21.7min$
\\ \hline
\end{tabular}
\caption{The computational time of the hybrid collision model with different $\theta_n$.}
\label{table1}
\end{table}

\subsection{The \texorpdfstring{$\alpha$}{} particle transport module test for ICF application}
\begin{figure}
  \centering
  \includegraphics[width=0.8\textwidth]{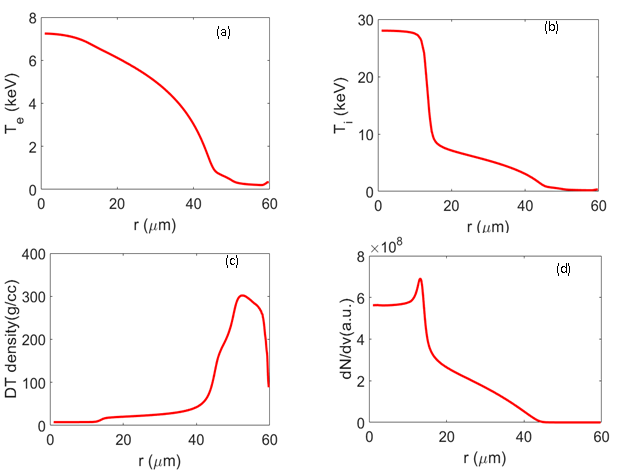}
  \caption{Plasma flow field of N170601 at bang time: 
  (a) electron temperature distribution, 
  (b) ion temperature distribution, 
  (c) D/T number density, 
  (d) alpha source term.}
  \label{fig_N170601}
\end{figure}

\begin{figure}
  \centering
  \includegraphics[width=0.45\textwidth]{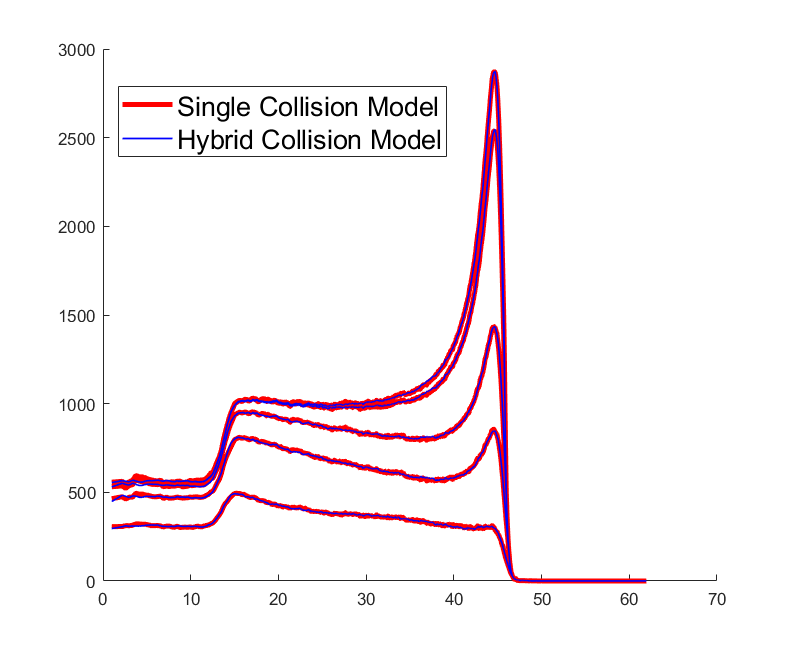}\\
  \includegraphics[width=0.43\textwidth]{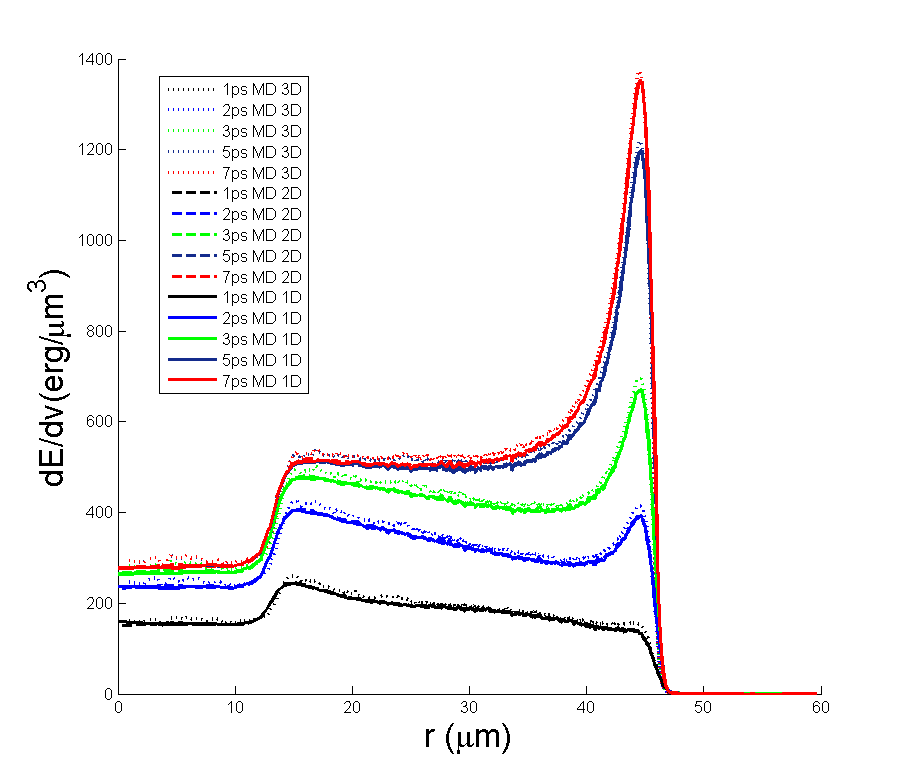}
  \includegraphics[width=0.43\textwidth]{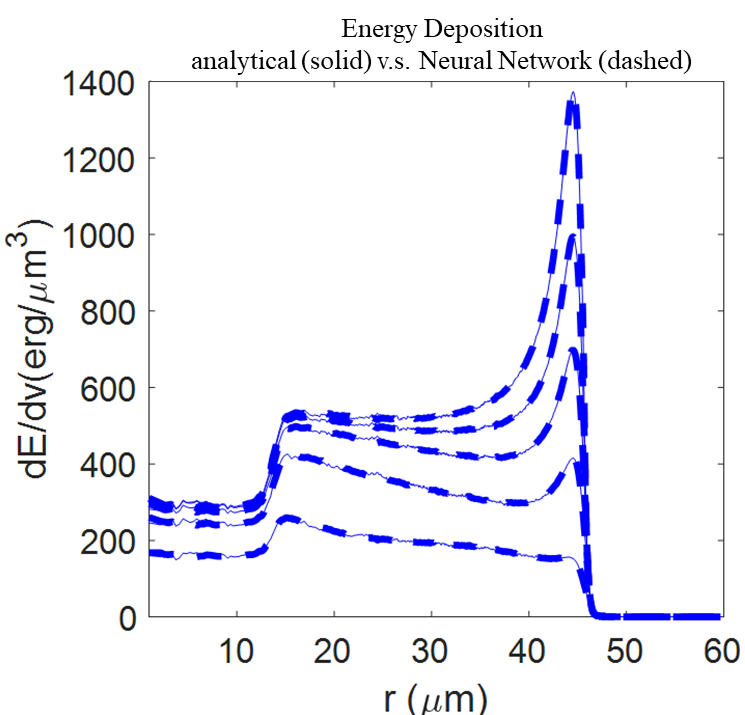}
  \caption{ The comparison of energy deposition profile in the radial direction between 
the traditional MC algorithm, the hybrid collision model, the 1-3 dimensional calculation, 
and the neural network results.}
  \label{fig_N170601energy}
\end{figure}

\begin{figure}
  \centering
  \includegraphics[width=0.4\textwidth]{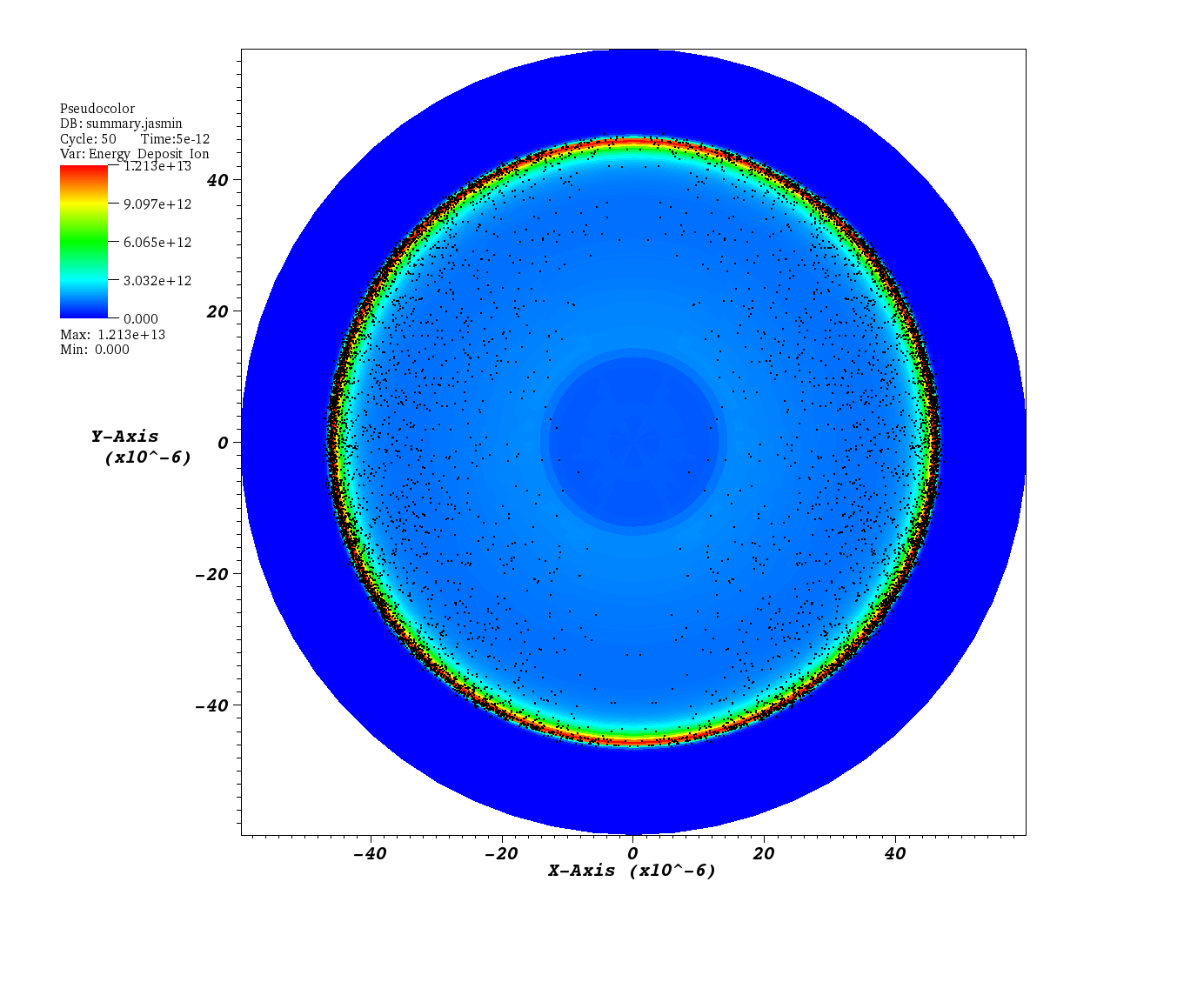}
  \includegraphics[width=0.4\textwidth]{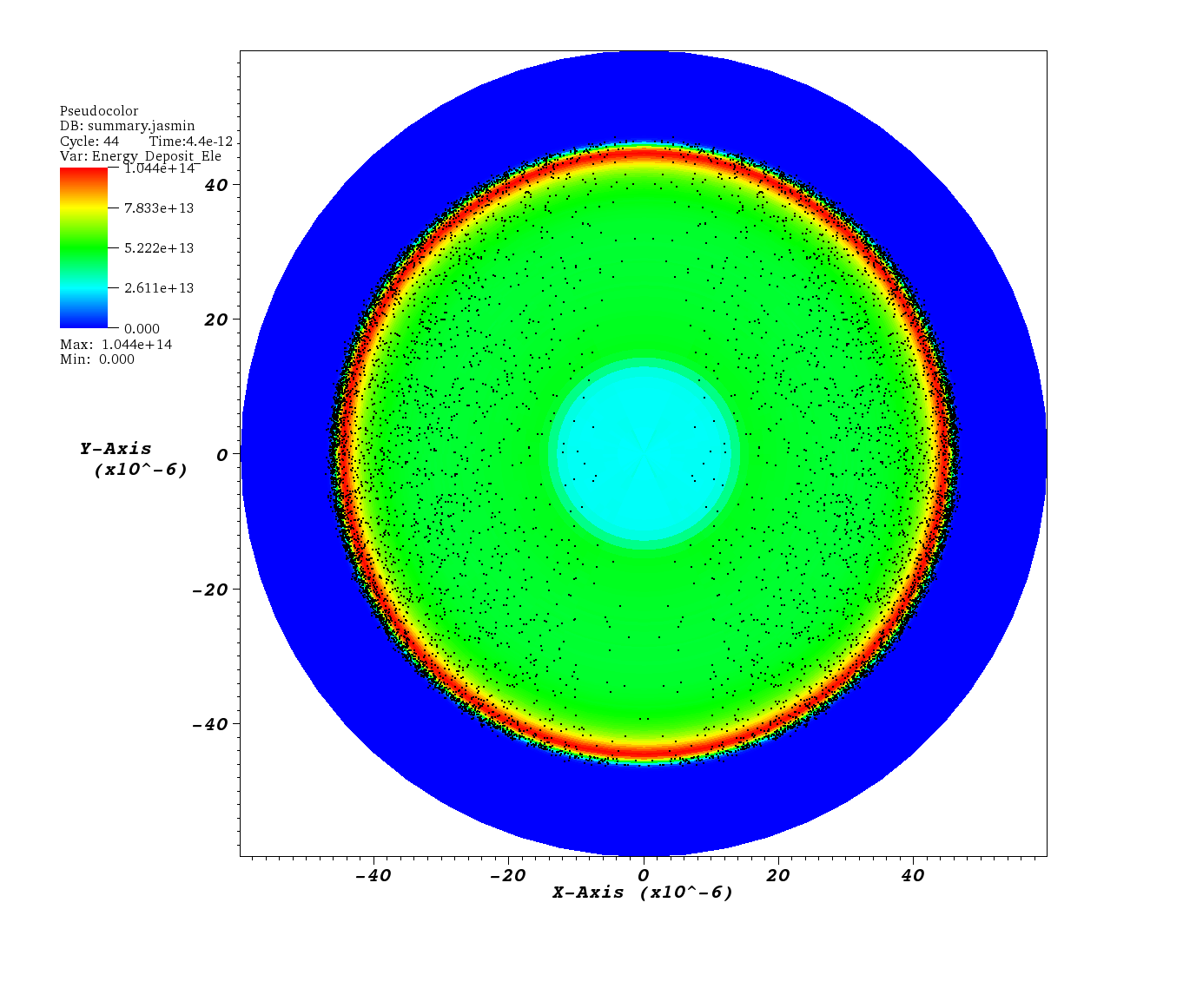}
  \caption{Two-dimensional charged particles transport energy deposition of N170601, the left figure is the ion energy deposition contour surface, and the right figure is the energy deposition contour surface in electrons.}
  \label{fig_N1706012d}
\end{figure}

\begin{figure}
  \centering
  \includegraphics[width=0.49\textwidth]{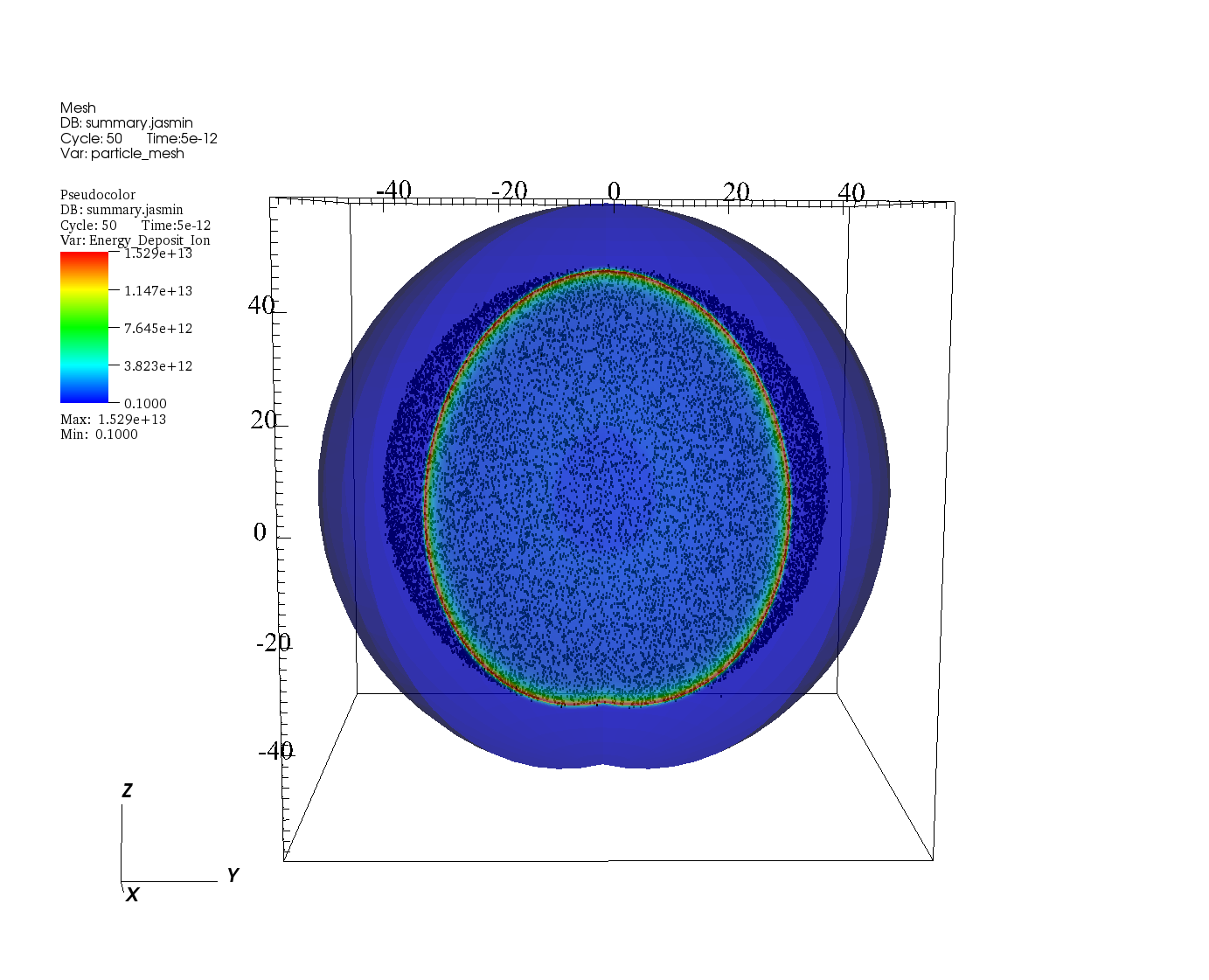}
  \includegraphics[width=0.49\textwidth]{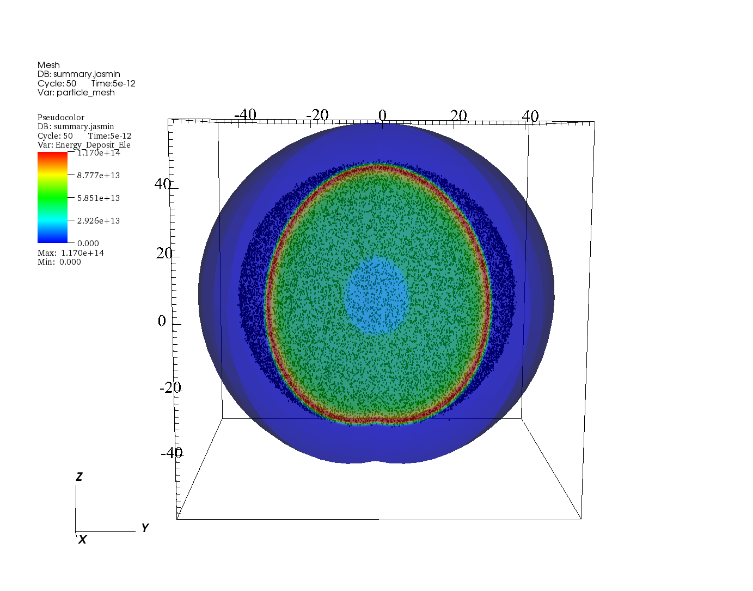}
  \caption{ Three-dimensional charged particles transport energy deposition of N170601, the left figure is the ion energy deposition contour surface, and the right figure is the energy deposition contour surface in electrons.}
  \label{fig_N1706013d}
\end{figure}

The $\alpha$ particle transport code module is tested by 
simulating the ICF N170601 experiment \cite{kritcher2022design}.
The distribution of the plasma flow field of 
N170601 at the bang time is shown in figure \ref{fig_N170601}.
We calculate the $\alpha$ energy deposition in $t=7\text{ ps}$.
The contour diagram of charged particle energy in ions and electrons 
for two-dimensional and three-dimensional geometry is shown 
in figure \ref{fig_N1706012d} and \ref{fig_N1706013d}.
The comparison of energy deposition profile in the radial direction between 
the traditional MC algorithm, the hybrid collision model, the 1-3 dimensional calculation, 
and the neural network results are shown in figure \ref{fig_N170601energy}.
It is shown that the proposed method provides consistent results with the traditional methods, 
which verifies the $\alpha$ particle transport code module.


\subsection{Development of the integrated ICF software with the \texorpdfstring{$\alpha$}{} particle transport module}
\begin{figure}
  \centering
  \includegraphics[width=0.49\textwidth]{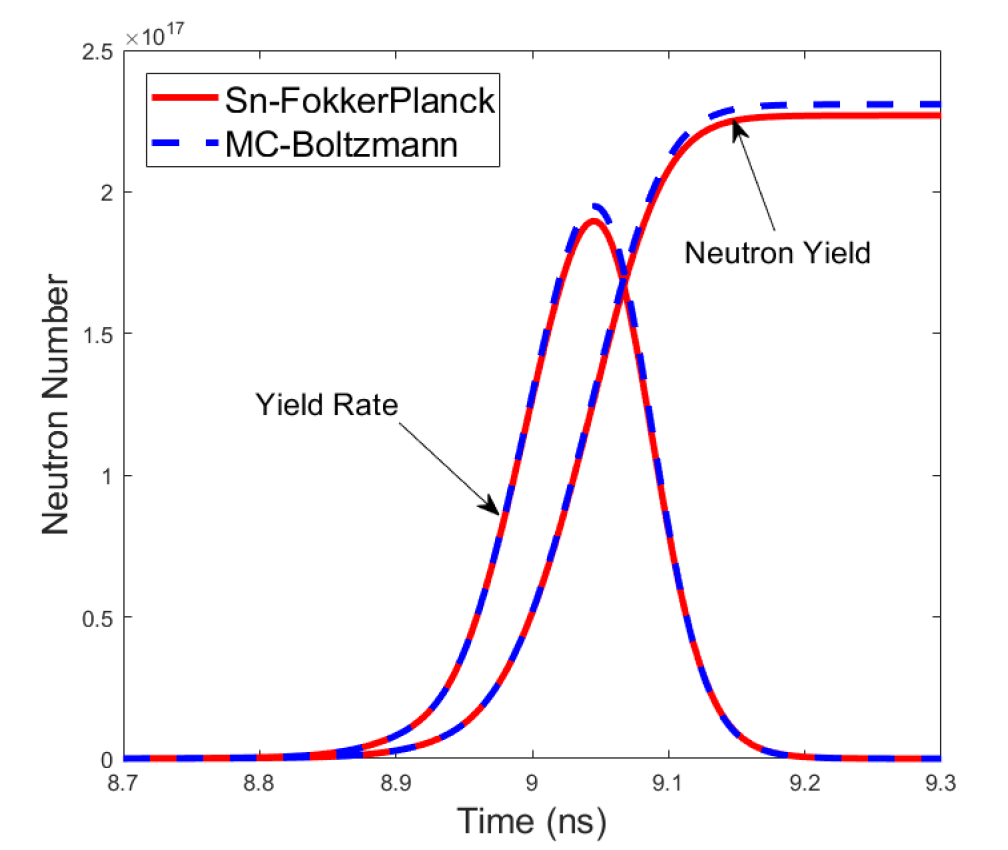}
  \includegraphics[width=0.49\textwidth]{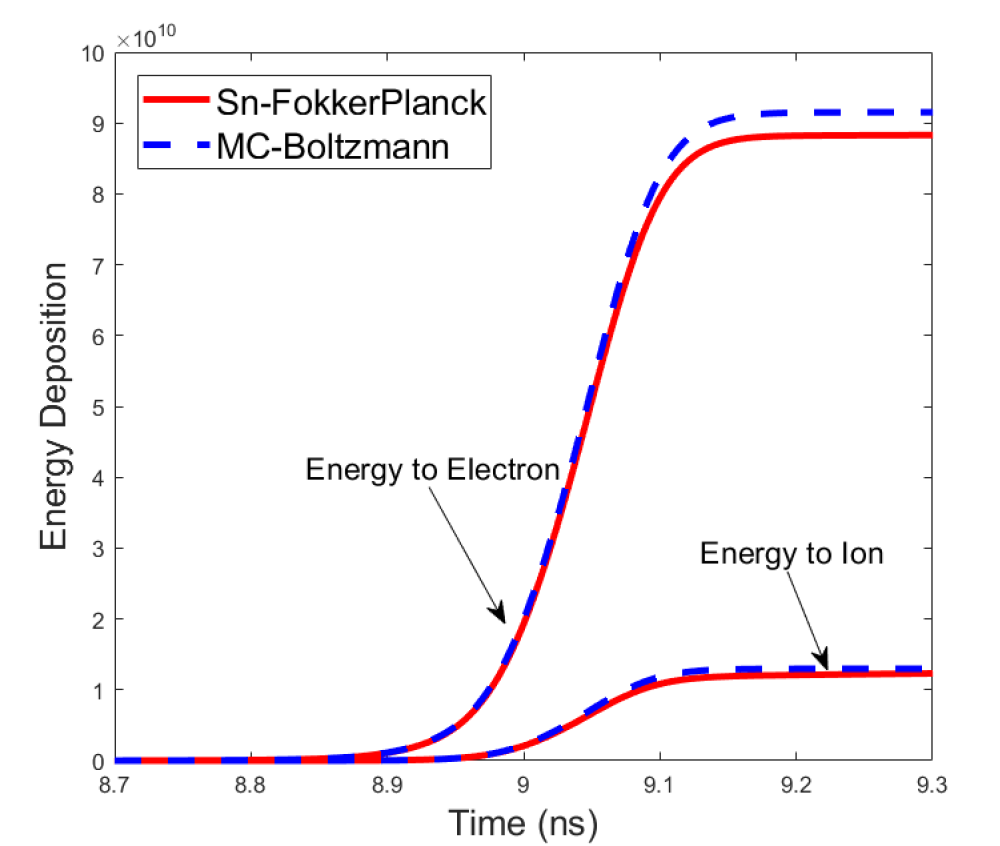}
  \caption{The simulation of N191110 experiment. 
  The neutron yield and neutron production rate are shown in the left figure, and 
  the charged particle energy deposition in electrons and ions is shown in the right figure. 
  The dashed line is the current hybrid-MC results, and the solid line is the SN results.}
  \label{fig_N1911101}
\end{figure}

\begin{figure}
  \centering
  \includegraphics[width=0.45\textwidth]{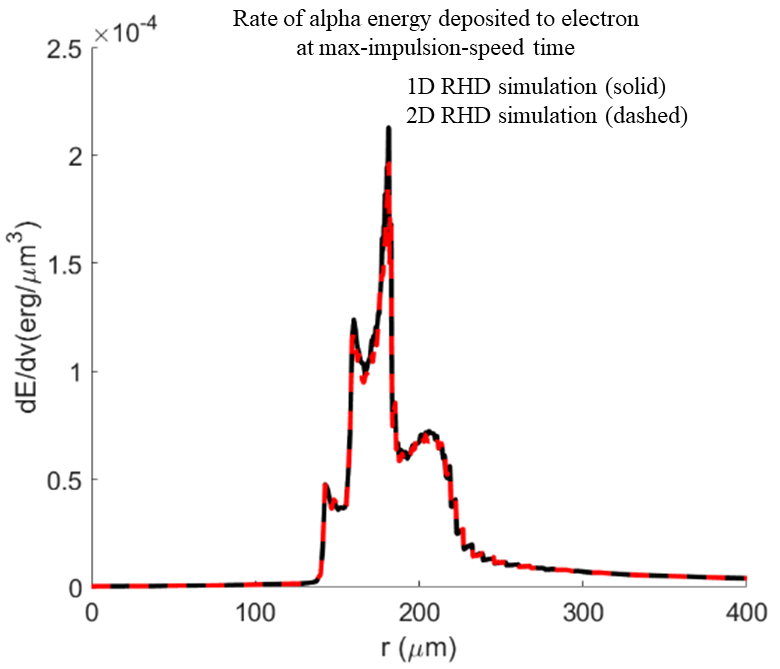}
  \includegraphics[width=0.45\textwidth]{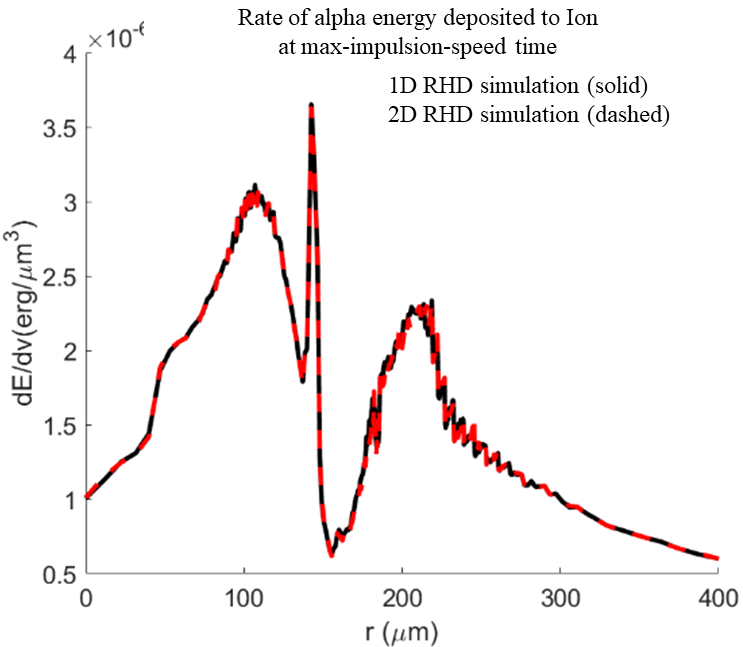}
  \caption{ Comparison of $\alpha$ particle energy deposition at the maximum-implosion-velocity time. 
  The energy deposition of $\alpha$ particles in electrons is shown in the left figure, and 
  the energy deposition of $\alpha$ particles in ions is shown in the right figure. 
  The solid black line is the 1D RDMG result, and the dotted red line is the 2D LARED-S program result.}
  \label{fig_N1911102}
\end{figure}

\begin{figure}
  \centering
  \includegraphics[width=0.45\textwidth]{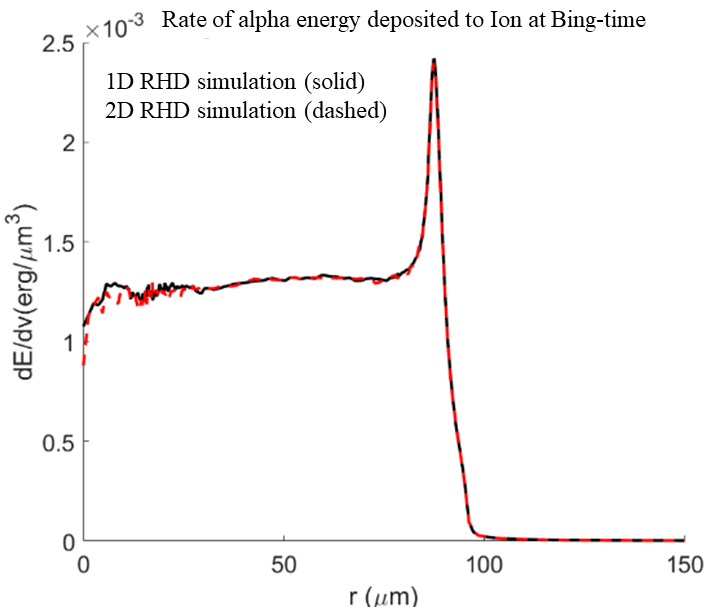}
  \includegraphics[width=0.45\textwidth]{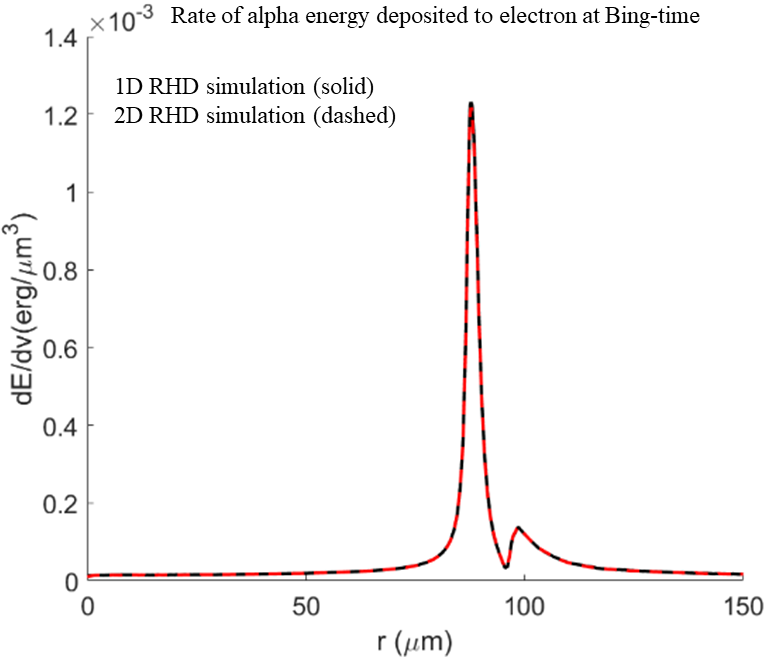}
  \caption{The comparison of $\alpha$ particle energy deposition results at Bang time. 
  The energy deposition of $\alpha$ particles in electrons is shown in the left figure, and 
  the energy deposition of $\alpha$ particles in ions is shown in the right figure. 
  The solid black line is the 1D RDMG result, and the dotted red line is the 2D LARED-S program result.}
  \label{fig_N1911103}
\end{figure}

\begin{figure}
  \centering
  \includegraphics[width=0.45\textwidth]{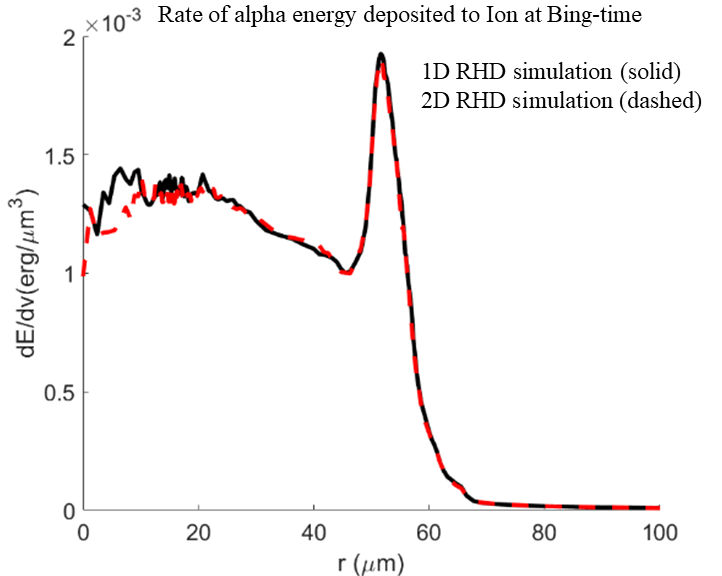}
  \includegraphics[width=0.45\textwidth]{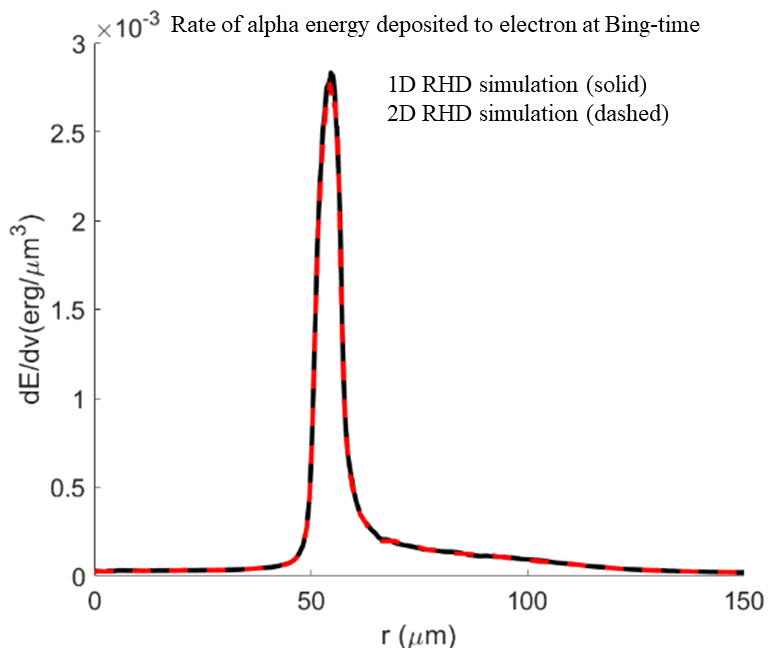}
  \caption{ The comparison of $\alpha$ particle energy deposition results at the stagnation time.
  The energy deposition of $\alpha$ particles in electrons is shown in the left figure, and 
  the energy deposition of $\alpha$ particles in ions is shown in the right figure. 
  The solid black line is the 1D RDMG result, and the dotted red line is the 2D LARED-S program result.}
  \label{fig_N1911104}
\end{figure}

\begin{figure}
  \centering
  \includegraphics[width=0.52\textwidth]{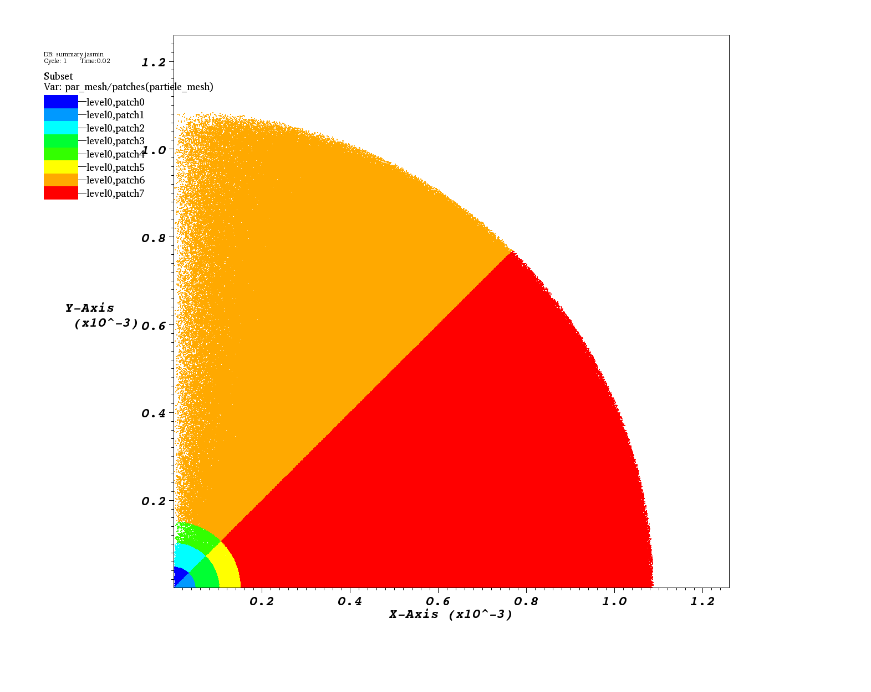}
  \includegraphics[width=0.47\textwidth]{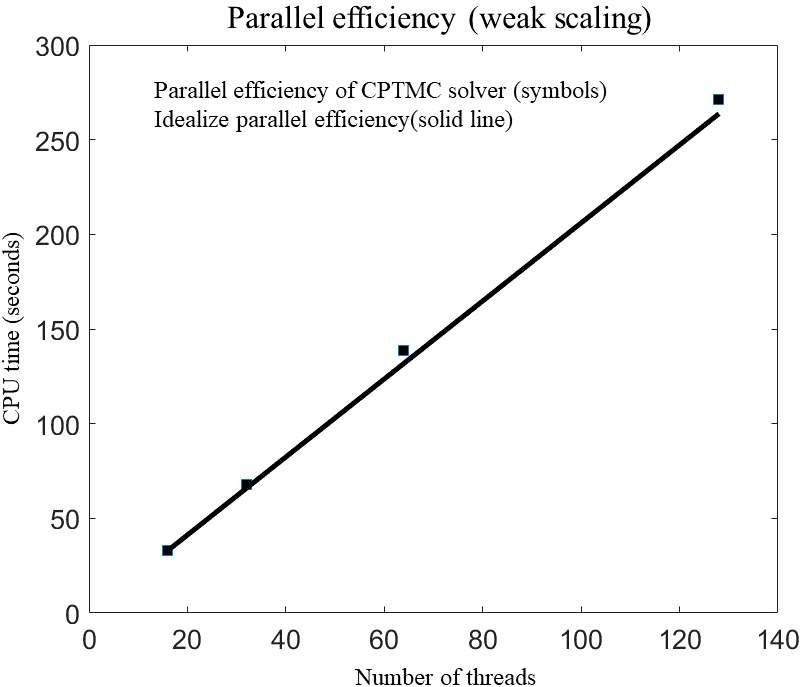}
  \caption{The parallel zones and parallel efficiency of the 2D LARED-S program.}
  \label{fig_parallel}
\end{figure}

We develop the one-dimensional RDMG code and the two-dimensional LARED-S code for an integrated ICF simulation \cite{pei2007construction}. 
Both codes have been applied in the simulation of the ICF N191110 experiment \cite{kritcher2022design}.
A comparison between the hybrid MC version code and the SN version code is shown in figure \ref{fig_N1911101}.
The error of neutron reaction rate and energy deposition is around $5\%$. 
The calculation time of the SN version was 5080s, and that of the MC version was 5084s, which is comparable.
The comparison between the 1D and 2D simulation flow field at 
the maximum-implosion-velocity time, the stagnation time, and the bang time 
are shown in figure \ref{fig_N1911102}-\ref{fig_N1911104}, 
A good agreement is observed, which shows the accuracy and robustness of the multi-dimensional code.
The parallel zones and parallel efficiency of the 2D LARED-S program are shown in figure \ref{fig_parallel},
which shows good parallel efficiency for small-scale simulations.


\section{Conclusion}\label{section_conclusion}
In this work, we develop an efficient numerical simulation method for 
$\alpha$ particle transport based on a hybrid collision model and machine learning.
The hybrid collision model improves the computational efficiency of the $\alpha$ particle transport 
by two orders of magnitude.
The machine learning based stopping power neutral network provides 
an extendable and effective algorithm for 
the calculation of $\alpha$ particle energy deposition.
The multi-dimensional $\alpha$ transport code modules are developed, and 
integrated into the multi-physics ICF software.
The accuracy and efficiency of the current MC version ICF software is 
verified by a simulation study of the N191110 experiment.

\section*{Acknowledgement}
The authors are partially supported by the National Natural Science Foundation of China (12102061) and 
the National Key R\&D Program of China (2022YFA1004500).
Chang Liu is partially supported by 
the Foundation of President of China Academy of Engineering Physics (YZJJZQ2022017).
Peng Song is partially supported by the National Natural Science Foundation of China (12031001).

\bibliographystyle{unsrt}
\bibliography{cptmc}
\end{document}